\documentclass[a4paper,11pt,preprintnumbers]{article}
\pdfoutput=1

\usepackage{multirow}
\usepackage{amssymb,mathrsfs,amsmath}
\usepackage{a4wide}
\usepackage{color,xcolor}
\usepackage{slashed,soul}
\usepackage{graphicx}
\usepackage{amsfonts}
\usepackage{cancel}
\usepackage{lscape}
\def\linkcolor{cyan!70!black}

\usepackage[
colorlinks=true
,urlcolor=\linkcolor
,anchorcolor=\linkcolor
,citecolor=\linkcolor
,filecolor=\linkcolor
,linkcolor=\linkcolor
,menucolor=\linkcolor
,linktocpage=true
,pdfproducer=medialab
,pdfa=true
]{hyperref}
\usepackage{amsthm}
\usepackage{booktabs}
\usepackage{array}
\usepackage{rotating}
\usepackage[numbers, sort&compress]{natbib}
\usepackage{multirow}
\usepackage{float}
\usepackage[utf8]{inputenc}
\usepackage[T1]{fontenc}
\usepackage{appendix}
\usepackage{epsfig}
\usepackage{orcidlink}
\usepackage{comment}
\usepackage{tabularx}

\newcommand{\beq}{\begin{equation}} 
\newcommand{\eeq}{\end{equation}} 
\newcommand{\ba}{\begin{array}}  
\newcommand{\ea}{\end{array}} 
\newcommand{\bea}{\begin{eqnarray}}  
\newcommand{\eea}{\end{eqnarray} }  
\newcommand{\bal}{\begin{align}}
\newcommand{\eal}{\end{align}}   
\newcommand{\bi}{\begin{itemize}}  
\newcommand{\ei}{\end{itemize}}  
\newcommand{\ben}{\begin{enumerate}}  
\newcommand{\een}{\end{enumerate}}  
\newcommand{\bc}{\begin{center}}
\newcommand{\ec}{\end{center}} 
\newcommand{\bt}{\begin{table}}
\newcommand{\et}{\end{table}}  
\newcommand{\btb}{\begin{tabular}}
\newcommand{\etb}{\end{tabular}}

\def\arrvline{\hfil\kern\arraycolsep\vline\kern-\arraycolsep\hfilneg}

\newcolumntype{Y}{>{\centering\arraybackslash}X}

\definecolor{mycoral}{HTML}{FF7F50}

\allowdisplaybreaks

\let\OLDthebibliography\thebibliography
\renewcommand\thebibliography[1]{
  \OLDthebibliography{#1}
  \setlength{\parskip}{0pt}
  \setlength{\itemsep}{0pt plus 0.3ex}
}

\begin{document}

\vspace{1cm}

\begin{titlepage}

\begin{flushright}
FTUV-26-0415.0632
 \end{flushright}
\vspace{0.2truecm}

\begin{center}
\renewcommand{\baselinestretch}{1.8}\normalsize
\boldmath
{\LARGE\textbf{
Sharpening New Physics Searches in Neutrino Oscillations with DUNE-PRISM}}
\unboldmath
\end{center}

\vspace{0.4truecm}

\renewcommand*{\thefootnote}{\fnsymbol{footnote}}

\begin{center}

{

Josu Hern\'andez-Garc\'ia$^1$\footnote{\href{mailto:josu.hernandez@ific.uv.es}{josu.hernandez@ific.uv.es}}\orcidlink{0000-0003-0734-0879},
Jacobo L\'opez-Pav\'on$^1$\footnote{\href{mailto:jacobo.lopez@ific.uv.es}{jacobo.lopez@ific.uv.es}}\orcidlink{0000-0002-9554-5075}

and Salvador Urrea$^{2}$\footnote{\href{mailto:salvador.urrea@ijclab.in2p3.fr}{salvador.urrea@ijclab.in2p3.fr}}\orcidlink{0000-0002-7670-232X}
}

\vspace{0.7truecm}

{\footnotesize
$^1$ Instituto de F\'{\i}sica Corpuscular, Universidad de Valencia and CSIC\\ 
 Edificio Institutos Investigaci\'on, Catedr\'atico Jos\'e Beltr\'an 2, 46980 Spain\\
$^2$ IJCLab, Pôle Théorie (Bat. 210), CNRS/IN2P3, 91405 Orsay, France
}

\vspace*{2mm}
%\today
\end{center}

\renewcommand*{\thefootnote}{\arabic{footnote}}
\setcounter{footnote}{0}

%\vspace{0.3cm}
\begin{abstract}
Upcoming long-baseline neutrino oscillation experiments such as DUNE aim to achieve unprecedented precision, but their physics reach is ultimately constrained by systematic uncertainties in neutrino flux predictions and neutrino–nucleus cross sections. These limitations are especially critical for new-physics searches in neutrino oscillations at the near detector, including non-unitarity and sterile neutrinos, where the signal manifests as small distortions in the energy spectrum and is therefore highly sensitive to spectral uncertainties. The PRISM (Precision Reaction Independent Spectrum Measurement) technique offers a robust strategy to mitigate these effects by exploiting measurements at multiple off-axis angles, effectively providing a data-driven handle to reduce systematics. In this work, we demonstrate that PRISM can significantly reduce the impact of large systematic uncertainties, restoring sensitivity to non-unitarity and sterile neutrino scenarios in the electron and muon sectors to a level comparable to that obtained with small spectral uncertainties. We also include the results for the $\tau$ sector with PRISM; however, in this case, since the majority of the flux measured at off-axis angles lies below the $\tau$ production threshold, we find the improvement to be marginal. As part of this work, we have obtained neutrino and antineutrino fluxes for different off-axis angles with higher statistics than those provided by the DUNE collaboration. We make available these fluxes as auxiliary material to this manuscript.

\end{abstract}

\end{titlepage}

\tableofcontents

%%%%%%%%%%%%%%%%%%%%%%%%%%%
\section{Introduction}

The observation of neutrino oscillations has firmly established that at least two neutrino species are massive and that leptonic flavors mix~\cite{SNO:2002tuh,Super-Kamiokande:1998kpq,KamLAND:2002uet}. However, the origin of neutrino masses remains unknown and requires physics beyond the Standard Model. A broad class of extensions introduces additional neutral fermions that mix with the active neutrinos, generating the tiny neutrino masses either through mixing with much heavier states, as in seesaw mechanisms~\cite{Minkowski:1977sc, Mohapatra:1979ia, Yanagida:1979as, GellMann:1980vs}, or via the additional suppression associated to an approximate symmetry such as lepton number~\cite{Branco:1988ex,Kersten:2007vk,Abada:2007ux,Moffat:2017feq}, as in low-scale realizations like the inverse seesaw~\cite{Mohapatra:1986bd,Gonzalez-Garcia:1988okv,Gavela:2009cd,BERNABEU1987303,Mohapatra:1986aw} or linear seesaw~\cite{Akhmedov:1995ip, Barr:2003nn, Malinsky:2005bi}. In general, these scenarios lead to an enlarged leptonic mixing matrix, where the effective $3\times 3$ active sub-block is no longer unitary. The phenomenology depends strongly on the mass scale of the new states. For heavy masses above the electroweak scale ($m \gtrsim m_{\rm EW}$), the new states induce non-unitarity effects that modify electroweak and flavor observables, which constrain deviations from unitarity at the per-mille level~\cite{Blennow:2023mqx}. For much lighter states, $10\,\mathrm{eV} \lesssim m \lesssim 1\,\mathrm{MeV}$, they can be treated as effectively massless at high energies, such that the standard predictions for precision observables are recovered. In this regime, the dominant constraints arise from neutrino oscillation experiments. In this intermediate mass range, the heavy states can be kinematically produced in neutrino experiments, but the oscillations they induce are too fast to be resolved even at very short baselines. Their effect is analogous to an effective non-unitarity of the $3\times 3$ active mixing matrix~\cite{Blennow:2016jkn}, a scenario commonly referred to as low-scale non-unitarity. For even lighter states, the new degrees of freedom can actively participate in oscillations, leading to additional oscillation frequencies typically associated with sterile neutrinos, which have historically been invoked to explain anomalies such as those observed by LSND~\cite{Aguilar:2001ty} and MiniBooNE~\cite{Aguilar-Arevalo:2020nvw}.

Neutrino oscillation physics is entering a precision era. With high-power accelerator beams, large-mass detectors, and steadily improving reconstruction and analysis methods, statistical uncertainties are no longer the dominant limitation for many key measurements. Instead, the ultimate reach of next-generation long-baseline programs such as DUNE~\cite{DUNE:2020ypp,DUNE:2020jqi} and Hyper-Kamiokande~\cite{Abe:2015zbg,Abe:2018uyc} will be determined by systematic uncertainties, most notably those originating from hadron production, which drive neutrino flux uncertainties, and from neutrino--nucleus interaction modeling in the few-GeV regime~\cite{NuSTEC:2017hzk}.

To mitigate these systematics in standard oscillation measurements, both experiments rely on near detector (ND) complexes. Located close to the source, these detectors provide high-statistics measurements of the unoscillated neutrino beam and allow for in situ constraints on flux and cross-section uncertainties, which would otherwise propagate to the far-detector (FD) analysis~\cite{DUNE:2021cuw, Hyper-Kamiokande:2018ofw}. At the same time, the large statistics available at the ND make it an attractive environment for short-baseline searches of new physics~\cite{Antusch:2006vwa, Miranda:2018yym,Coloma:2021uhq}. However, this opportunity comes with an important caveat: since the ND is primarily designed to control systematics for the FD, it is itself affected by large uncertainties, both in normalization and in spectral shape. In many new-physics scenarios affecting neutrino oscillations such as non-unitarity and sterile neutrinos, current bounds constrain deviations from the standard picture to be small. As a result, and given that the normalization uncertainties associated with fluxes and cross sections are larger than the current bounds, the sensitivity at the ND is not driven by overall rate changes but rather by the precise knowledge of the energy spectrum. In other words, the ability to detect (or constrain) new physics crucially depends on controlling spectral-shape uncertainties, which remain large and constitute the main limiting factor in these searches.

A promising handle on spectral uncertainties is provided by the PRISM (Precision Reaction Independent Spectrum Measurement) strategy~\cite{nuPRISM:2014mzw,Hasnip:2023ygr,Gehrlein:2025lbj,DUNE:2025lvs}. In this approach, the near detector samples the neutrino beam at multiple off-axis angles, accessing different energy spectra that are primarily determined by the decay kinematics of the parent mesons. Since neutrino interaction cross sections are common across these configurations, comparing measurements at different angles allows for a partial disentanglement of flux and interaction effects. Moreover, the relative differences between off-axis spectra are expected to be better controlled, as they are largely governed by well-understood decay kinematics and by horn focusing effects, whose associated uncertainties are smaller than those arising from hadron production and neutrino cross-section modeling.

While the PRISM strategy has been primarily discussed in the context of improving standard oscillation measurements, its potential for searches of new physics at the near detector has received comparatively less attention. In particular, the extent to which PRISM measurements can mitigate the impact of large spectral uncertainties in short-baseline searches remains to be fully quantified. In this work, we study the impact of PRISM on future DUNE searches for non-unitarity and sterile neutrinos. We assume a conservative treatment of spectral uncertainties, corresponding to a $5\%$ bin-to-bin uncorrelated shape uncertainty, which would dramatically reduce the sensitivity to these scenarios and show that measurements with PRISM allows to recover most of the lost sensitivity, yielding results comparable to those obtained with much smaller spectral uncertainties.

The remainder of this paper is organized as follows. In Section~\ref{sec:theory} we introduce the theoretical framework for non-unitarity and sterile neutrinos. In Section~\ref{sec:set-up} we describe the experimental configuration and simulation details of the DUNE near detector and the DUNE-PRISM program, including the computation of the off-axis neutrino fluxes, the treatment of systematic uncertainties, and the statistical methodology adopted in our analysis. Our main results are presented in Section~\ref{sec:results}, while the conclusions are summarized in Section~\ref{sec:summary}. The analysis and results for the $\nu_\tau$ sector are provided in Appendix~\ref{app:tau_detection}.

\section{Theoretical framework}\label{sec:theory}

Among the possible extensions of the SM that explain neutrino oscillation data and the smallness of neutrino masses, the simplest and most economical consists of adding singlet fermions to the SM field content. These new states naturally generate light neutrino masses either via suppression by a heavy scale, as in the type-I seesaw~\cite{Minkowski:1977sc,Yanagida:1979as,Yanagida:1979gs,Gell-Mann:1979vob}, or via the additional suppression given by an approximate symmetry such as lepton number, as in low-scale seesaw scenarios like the inverse and linear seesaws~\cite{Mohapatra:1986aw,Mohapatra:1986bd,Bernabeu:1987gr,Branco:1988ex,Malinsky:2005bi}. See Ref.~\cite{Cai:2017jrq,King:2003jb,Boucenna:2014zba} for reviews and more complete list of neutrino mass generation mechanisms. Within this broad class of neutrino mass models, the masses of the extra fermions can range from $\mathcal{O}({\rm eV})$ up to energies near the GUT scale, and these extra fermions—if allowed by the quantum numbers of the model—can mix with the light neutrinos, yielding an enlarged mixing matrix. Historically, the idea of additional sterile neutrinos with masses around the $\mathcal{O}({\rm eV})$ scale has been extensively explored, as they have been invoked as a possible explanation for short-baseline anomalies as the ones reported by LSND~\cite{Aguilar:2001ty}, MiniBooNE~\cite{Aguilar-Arevalo:2020nvw}, and the now alleviated reactor anomaly~\cite{Mention:2011rk,Huber:2011wv}.

Without resorting to any particular model, one may assume in full generality that $n$ extra fermions mix with the light neutrinos. The leptonic mixing matrix $\mathcal{U}$ that diagonalizes the full neutrino mass matrix and connects flavor and mass eigenstates is then a unitary $(3+n)\times(3+n)$ matrix, and it can be written in block form as
\begin{equation}
\mathcal{U} = \begin{pmatrix}
N & \Theta \\
R & S
\end{pmatrix} \, ,
\end{equation}
where $N$ denotes the $3\times 3$ active--light sub-block (which will play the role of an effective PMNS matrix) and will be in general non-unitary ($NN^\dagger \neq \mathbb{I}$). On the other hand, $\Theta$ %and $\mathcal{R}$ 
encode the mixing between the active neutrinos and the mostly sterile mass eigenstates that govern experimental signatures of heavy neutrinos.

%It is often the case that deviations from unitarity of the matrix $N$ are much better directly constrained than the individual $N$ matrix elements
Given that the bounds on the deviations from unitarity of the matrix $N$ are more stringent than the ones directly derived on the $N$ matrix elements~\cite{Blennow:2025qgd}, it is %therfore 
convenient to parametrize $N$ in a way that makes the deviation from unitarity explicit. Without loss of generality, the matrix $N$ can be parametrized in two equivalent ways commonly used in the literature:
\begin{equation}
N = (I - T)U \quad \mathrm{or} \quad N = (I - \eta)U',
\label{eq:N}
\end{equation}
where $\eta$ is a Hermitian matrix~\cite{Broncano:2002rw,FernandezMartinez:2007ms}\footnote{There is no loss of generality since using the polar decomposition, any square matrix can be written as the product of a Hermitian and a unitary matrix.}, while $(I - T)$ is a lower triangular matrix given by
\begin{equation}
T = \begin{pmatrix}
\alpha_{ee} & 0 & 0\\
\alpha_{\mu e} & \alpha_{\mu \mu} & 0\\
\alpha_{\tau e} & \alpha_{\tau \mu} & \alpha_{\tau \tau}
\end{pmatrix},
\label{eq:alpha}
\end{equation}
$U$ and $U'$ are unitary matrices that coincide with the standard PMNS matrix up to small corrections controlled by $\alpha$ and $\eta$. The two parameterizations are fully equivalent, and an explicit mapping between them is given in~\cite{Blennow:2016jkn}. In this work, we adopt the lower triangular parametrization, with results that can be straightforwardly translated into the hermitian parametrization.
Exploiting the unitarity of $\mathcal{U}$, one can relate the $N$ and $\Theta$ subblocks through
\begin{equation}
NN^\dagger + \Theta \Theta^\dagger = I.
\end{equation}
Introducing the triangular parametrization for $N$ defined in Eqs.~(\ref{eq:N}) and (\ref{eq:alpha}), and expanding to leading order in $\alpha$, this condition becomes
\begin{equation}
I = I - T - T^\dagger + \Theta \Theta^\dagger + \mathcal{O}(\alpha^2),
\end{equation}
from this relation, one obtains
\begin{equation}
\label{eq:alpha_mixing}
\alpha_{\beta\beta}
=
\frac{1}{2}\left(\Theta\Theta^\dagger\right)_{\beta\beta}
=
\frac{1}{2}\sum_{i=4}^n |\mathcal{U}_{\beta i}|^2 \,,
\end{equation}
\begin{equation}
\label{eq:alphavsU}
\alpha_{\gamma\beta}
=
\left(\Theta\Theta^\dagger\right)_{\gamma\beta}
=
\sum_{i=4}^n \mathcal{U}_{\gamma i}\mathcal{U}_{\beta i}^* \,.
\end{equation}
It is important to emphasize that these relations do not uniquely determine $\Theta$ in terms of the $\alpha$ parameters, but rather constrain only the Hermitian combination $\Theta \Theta^\dagger$.

Up to now, we have remained completely agnostic about the masses of the heavy neutrinos, however the location of this mass scale will determine the phenomenology. We therefore distinguish between:
\begin{itemize}
    \item \textbf{High-scale non-unitarity} ($m > m_{\rm EW}$): If the deviations from unitarity originate from very heavy states, integrating them out induces modifications to the charged-current and neutral-current couplings of the active neutrinos. In this regime, very stringent constraints arise from precision electroweak and flavor observables~\cite{Langacker:1988ur,Nardi:1994iv,Tommasini:1995ii,Antusch:2006vwa,Alonso:2012ji,Akhmedov:2013hec,Antusch:2014woa,Fernandez-Martinez:2015hxa,Fernandez-Martinez:2016lgt}.
    
    \item \textbf{Low-scale non-unitarity} ($100\,\mathrm{eV} < m < 1\,\mathrm{MeV}$)\footnote{We leave aside the intermediate regime $m_{\rm EW} > m \gtrsim 1\,\mathrm{MeV}$, where additional constraints, that are stronger than those from oscillations, become relevant.}: When the new states are light enough to be kinematically produced and effectively massless at the EW scale, unitarity is effectively restored, and the previously mentioned constraints from precision electroweak and flavor observables do not apply. In this case, the current bounds are driven by neutrino oscillation experiments which are significantly less stringent~\cite{Blennow:2025qgd}.
\end{itemize}

Neutrino oscillation experiments, both current and near-future, are not expected to provide leading constraints on high-scale non-unitarity. Therefore, in the remainder of this work we assume the low-scale non-unitarity scenario. Nevertheless, we will present our results in the $\alpha$ parametrization, which is valid in both cases, and explain how to translate them to the high-scale non-unitarity case.

\subsection{Low-scale non-unitarity (averaged-out sterile neutrino oscillations)}

As discussed above, we define the low-scale non-unitarity scenario as one in which all heavy neutrinos have masses in the range $100\,\mathrm{eV} < m < 1\,\mathrm{MeV}$. In this regime, the heavy states can be kinematically produced in the neutrino source as they are effectively massless for the production processes. 
However, they are assumed to be sufficiently heavy to induce oscillations that cannot be resolved, even at short-baseline experiments or near detectors. Under these assumptions, the oscillation probability $P_{\gamma\beta}$ at a distance close to the source is given by
\begin{equation}
P_{\gamma \beta}
=
\left|\left(\mathcal U\,\mathcal S\,\mathcal U^\dagger\right)_{\beta\gamma}\right|^2 \,,
\end{equation}
where $\mathcal S$ is the vacuum evolution matrix,
\begin{equation}
\mathcal{S}=\operatorname{diag}\left(\exp \left(-i  \frac{\Delta m_{j 1}^2L}{2 E}\right)\right).
\end{equation}

At short baselines, the phases associated with the light states satisfy $\Delta m_{j1}^2 L / (2E) \ll 1$ for $j=1,2,3$, while for the heavy states one has $\Delta m_{j1}^2 L / (2E) \gg 1$ for $j \geq 4$. As a result, the oscillations induced by the heavy states are too rapid to be resolved and must be averaged over the detector energy resolution. This leads to the following averaged probability
\begin{equation}
\langle P_{\gamma\beta}\rangle
=
\left| \left(NN^\dagger\right)_{\beta\gamma}\right|^2
+
\sum_{I=4}^{n}
|\Theta_{\beta I}|^2\,|\Theta_{\gamma I}|^2 \,.
\end{equation}

It is useful to express this result in terms of the triangular parametrization ($\alpha$). For appearance channels ($\gamma \neq \beta$), one obtains
\begin{equation}
\label{eq:Pab_alphatheta}
\langle P_{\gamma\beta}\rangle
=
\left|\alpha_{\beta \gamma} \right|^2
+
\sum_{I=4}^{n}
|\Theta_{\beta I}|^2\,|\Theta_{\gamma I}|^2
+ \mathcal O(\alpha^3)\,,
\end{equation}
while for disappearance channels,
\begin{equation}
\label{eq:Paa_alphatheta}
\langle P_{\beta\beta}\rangle
=
1-4\alpha_{\beta\beta}
+
\sum_{I=4}^{n}
|\Theta_{\beta I}|^4
+ \mathcal O(\alpha^2)\,.
\end{equation}

In general, the second and third terms of Eq.~(\ref{eq:Pab_alphatheta}) and Eq.~(\ref{eq:Paa_alphatheta}) respectively cannot be expressed solely in terms of $\alpha$, see Eqs.~\eqref{eq:alpha_mixing} and \eqref{eq:alphavsU}. 
More explicitly, we can write the Hermitian combination in $\alpha_{\gamma\beta}$ as:
\begin{equation}
\label{eq:hermitian_combination}
\left|\left(\Theta \Theta^{\dagger}\right)_{\gamma \beta}\right|^2
=
\sum_{i=4}^n\left|\Theta_{\gamma i}\right|^2\left|\Theta_{\beta i}\right|^2
+
\sum_{\substack{i,j=4 \\ i \neq j}}^n
\Theta_{\gamma i} \Theta_{\beta i}^* \Theta_{\gamma j}^* \Theta_{\beta j} \,.
\end{equation}

A simplification occurs in the presence of a single heavy state. In that case, the interference term in Eq.~\eqref{eq:hermitian_combination} vanishes, yielding
\begin{equation}
\left|\left(\Theta \Theta^{\dagger}\right)_{\alpha \beta}\right|^2
=
\left|\Theta_{\alpha 4}\right|^2
\left|\Theta_{\beta 4}\right|^2 \,.
\end{equation}
The averaged probabilities can then be written entirely in terms of the $\alpha$ parametrization:
\begin{equation}
\label{eq:P_3plus1_app}
\langle P^{\text{3+1}}_{\gamma\beta}\rangle
=
2\left|\alpha_{\beta \gamma} \right|^2
+ \mathcal O(\alpha^3)\,,
\end{equation}
\begin{equation}
\label{eq:P_3plus1_diss}
\langle P^{\text{3+1}}_{\beta\beta}\rangle
=
1-4\alpha_{\beta\beta}
+ \mathcal O(\alpha^2)\,.
\end{equation}

More generally, independently of the number of heavy neutrino states, one can derive the inequalities
\begin{equation}
\label{eq:inequality_app}
\langle P_{\gamma\beta}\rangle
\geq
\left|\alpha_{\beta \gamma} \right|^2
+ \mathcal O(\alpha^3)\,,
\end{equation}
\begin{equation}
\label{eq:inequality_diss}
\langle P_{\beta\beta}\rangle
\geq
1-4\alpha_{\beta\beta}
+ \mathcal O(\alpha^2)\,.
\end{equation}

Therefore, adopting the equalities in Eqs.~\eqref{eq:inequality_app} and ~\eqref{eq:inequality_diss}  leads to conservative bounds that remain valid for an arbitrary number of heavy states. In the remainder of this work, we will use, however, the expressions in Eqs.~\eqref{eq:P_3plus1_app}~and~\eqref{eq:P_3plus1_diss} to facilitate comparison with our sterile neutrino analyses performed in the $3+1$ scenario. The current bounds in this case can be found in Table~\ref{tab:NU_averaged}, in order to obtain conservative bounds for any number of heavy species, the off-diagonal $\alpha$ parameters can be rescaled by a factor of $\sqrt{2}$.

\begin{table}[t]
    \centering
    \renewcommand{\arraystretch}{1.15}
    \begin{tabular}{lcc}
    \hline
     & \multicolumn{2}{c}{\bf Low--scale non--unitarity ($100\,\mathrm{eV} < m < 1\,\mathrm{keV}$)} \\
    \cline{2-3}
    {\bf 90\% CL} & {\bf Direct} & {\bf Schwarz} \\
    \hline
    $\alpha_{ee}$            & $8.4 \times 10^{-3}$ & -- \\
    $\alpha_{\mu\mu}$        & $1.2 \times 10^{-2}$ & -- \\
    $\alpha_{\tau\tau}$      & $2.9 \times 10^{-2}$ & -- \\
    $|\alpha_{\mu e}|$       & $1.6 \times 10^{-2}$ & $2.0 \times 10^{-2}$ \\
    $|\alpha_{\tau e}|$      & $6.1 \times 10^{-2}$ & $3.1 \times 10^{-2}$ \\
    $|\alpha_{\tau\mu}|$     & $9.1 \times 10^{-3}$ & $3.7 \times 10^{-2}$ \\
    \hline
    \end{tabular}
    \caption{Constraints on the non--unitarity parameters $\alpha_{\beta\gamma}$ from neutrino oscillation searches in the averaged--out regime. Stronger limits may apply for lighter sterile neutrinos where the oscillation pattern can be resolved. The values for the bounds has been taken from Ref.~\cite{Blennow:2025qgd} with the exception of the direct constraint in $|\alpha_{\mu e}|$ which has been updated with the recent results of MicroBooNE~\cite{MicroBooNE:2025nll}.}
    \label{tab:NU_averaged}
\end{table}

\subsection{Sterile neutrino oscillations}

In the previous discussion we assumed that the additional neutrino states are sufficiently heavy that their effects cannot be resolved, even at short baselines. We now instead examine the situation in which these extra states are light enough to actively participate in oscillations, leading to observable new oscillation signatures. For simplicity, we restrict ourselves to the minimal extension with a single additional state, the so-called $3+1$ sterile neutrino scenario.

In the short-baseline limit (relevant, for instance, to the DUNE near detector), the oscillations driven by the standard three neutrino mass splittings are not developed. As a result, the oscillation probabilities will only have contributions from the heavy neutrino state:
\begin{equation}
P_{\gamma\beta}
=
4\,|\mathcal{U}_{\beta 4}|^2 |\mathcal{U}_{\gamma 4}|^2
\sin^2\!\left(\frac{\Delta m^2_{41} L}{4E}\right),
\end{equation}

\begin{equation}
P_{\beta\beta}
=
1
-
4\,|\mathcal{U}_{\beta 4}|^2
\left(1 - |\mathcal{U}_{\beta 4}|^2\right)
\sin^2\!\left(\frac{\Delta m^2_{41} L}{4E}\right),
\label{eq:Psteriles}
\end{equation}
%

%It is often convenient 
In order to ease the comparison with most of the literature, it is convenient to rewrite these expressions in terms of effective mixing angles that directly characterize appearance and disappearance channels:
\begin{equation}
P_{\gamma\beta}
=
\sin^2 2\vartheta_{\gamma\beta}\,
\sin^2\!\left(\dfrac{\Delta m^2_{41} L}{4E}\right),
\quad
\text{with}
\quad
\sin^2 2\vartheta_{\gamma\beta}
\equiv
4\,|\mathcal{U}_{\beta 4}|^2 |\mathcal{U}_{\gamma 4}|^2,
\end{equation}

\begin{equation}
P_{\beta\beta}
=
1
-
\sin^2 2\vartheta_{\beta\beta}\,
\sin^2\!\left(\dfrac{\Delta m^2_{41} L}{4E}\right),
\quad
\text{with}
\quad
\sin^2 \vartheta_{\beta\beta}
\equiv
|\mathcal{U}_{\beta 4}|^2 \, .
\label{eq:sterile-eff-angles}
\end{equation}

Note that in the limit of fast oscillations $\sin^2\!\left(\dfrac{\Delta m^2_{41} L}{4E}\right) \to \cfrac{1}{2}$ and, using the relations in Eqs.~\eqref{eq:alpha_mixing} and \eqref{eq:alphavsU}, the above equations lead to the non-unitarity results discussed in Eqs.~\eqref{eq:P_3plus1_app} and \eqref{eq:P_3plus1_diss}.

\section{Experimental set-up and simulation details}
\label{sec:set-up}

\subsection{The DUNE-PRISM setup}

The Deep Underground Neutrino Experiment (DUNE) at Fermilab (USA) is a next-generation neutrino experiment aiming to shed light on some of the unknown properties of neutrinos in the context of particle physics, astrophysics, and cosmology.

At the core of its design is the high-intensity neutrino beam provided by the Long-Baseline Neutrino Facility (LBNF). The beam is generated by colliding 120~GeV protons against a 1.5~m-long segmented graphite target. The secondary particles produced in these collisions are subsequently focused by an optimized three-horn magnetic system and enter a 194~m-long decay volume where they propagate and decay into neutrinos. By reversing the current polarity, these horns focus or deflect the charged parent particles, selecting the operational mode of the experiment. The Forward Horn Current (FHC) setting produces a predominantly neutrino beam, while the Reverse Horn Current (RHC) setting produces a predominantly antineutrino beam. At the end of the decay pipe there is an absorber to stop the remaining secondary particles as well as the protons that passed through the target. A subsequent muon shield ensures a beam composed of neutrinos.

To measure neutrino oscillations with high precision, DUNE utilizes two primary detector complexes exposed to this beam. The Far Detector (FD), located 1300 km downstream at the Sanford Underground Research Facility (SURF), will consist of four 17~kton Liquid Argon Time Projection Chamber (LArTPC) modules. To characterize the unoscillated neutrino flux and constrain systematic uncertainties, the Near Detector (ND) complex is situated at Fermilab,  574~m downstream from the center of the target station. The ND hall is expected to %will 
host three detectors, the first being a liquid argon TPC (ND-LAr) sharing the same technology and nuclear target as the FD modules. ND-LAr features an active volume~\cite{DUNE:2020fgq} that is 7~m wide, 3~m high, and 5~m long, with a fiducial mass of 67.2~tons\footnote{In this work, we assume a fiducial volume excluding detector regions within 0.5~m of the walls. This conservative approach could be slightly refined by improving the event reconstruction in the outer regions of the active volume.}. Further downstream, there will be two secondary detectors: a muon spectrometer (TMS), and an on-axis magnetized neutrino detector (SAND). To enable the DUNE-PRISM experimental configuration, ND-LAr~$+$~TMS are mounted on a set of rails, allowing them to move perpendicular to the beam axis to perform measurements at various off-axis angles. DUNE Phase I is expected to start running in 2031, with a 1.2~MW beam delivering $1.1\cdot 10^{21}$ protons on target per year (POT/year). A possible Phase II is scheduled for 2035, with a beam upgrade to 2.4~MW delivering an average of $\sim 2.0\cdot 10^{21}$~(POT/year). This Phase II could also bring an upgrade to the ND, replacing TMS with a 1~ton high-pressure gaseous argon TPC detector (ND-GAr) surrounded by an electromagnetic calorimeter and a 0.5~T magnetic field. 

\subsection{Neutrino fluxes at DUNE-PRISM
}
\label{sec:neutrino_flux}

To model the expected neutrino beam, the DUNE collaboration utilizes a set of GEANT4-based simulations (G4LBNF)~\cite{DUNE:2021cuw,ntuples}. In its latest version, a full description of the target station, the horn system, and the beam absorber placed at the end of the decay volume have been implemented in order to simulate the production of light mesons with significant branching ratios to neutrinos.

In order to compute the neutrino fluxes at different off-axis locations, we have used the HNLux software developed in Ref~\cite{Coloma:2020lgy} to process the pool of parent mesons stored in the G4LBNF TTree tarballs. HNLux decays these mesons into light neutrinos, propagates them to the ND hall, and selects the neutrinos that intersect the ND-LAr geometry. Using this pipeline, the neutrino and antineutrino fluxes crossing ND-LAr are simulated for the standard DUNE on-axis position, as well as for seven distinct displacements of the detector corresponding to seven off-axis positions of the DUNE-PRISM experiment. Table~\ref{tab:off-axis} summarizes the off-axis angles and the corresponding ND-LAr displacements assumed for these simulations.

\begin{table}[h]
\centering
\begin{tabular}{|l|c|ccccccc|}
\hline
\textbf{Experiment} & \textbf{DUNE} & \multicolumn{7}{c|}{\textbf{PRISM}} \\ \hline
Off-axis angle [mrad] & 0 & 10.45 & 20.90 & 31.36 & 41.81 & 52.26 & 62.72 & 104.15 \\ \hline
Displacement [m] & 0 & 6.0 & 12.1 & 18.1 & 24.1 & 30.2 & 36.2 & 59.8 \\ \hline
\end{tabular}
\caption{DUNE-PRISM off-axis configurations assumed in our simulations.}
\label{tab:off-axis}
\end{table}

\begin{figure}[p]
    \centering
    \includegraphics[width=0.496\textwidth]{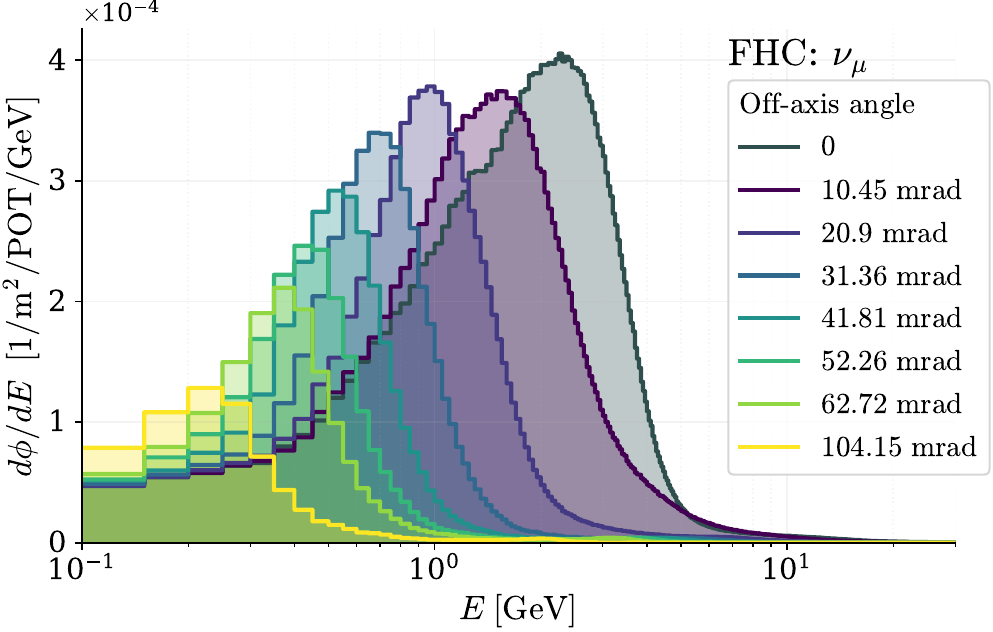}
    \includegraphics[width=0.496\textwidth]{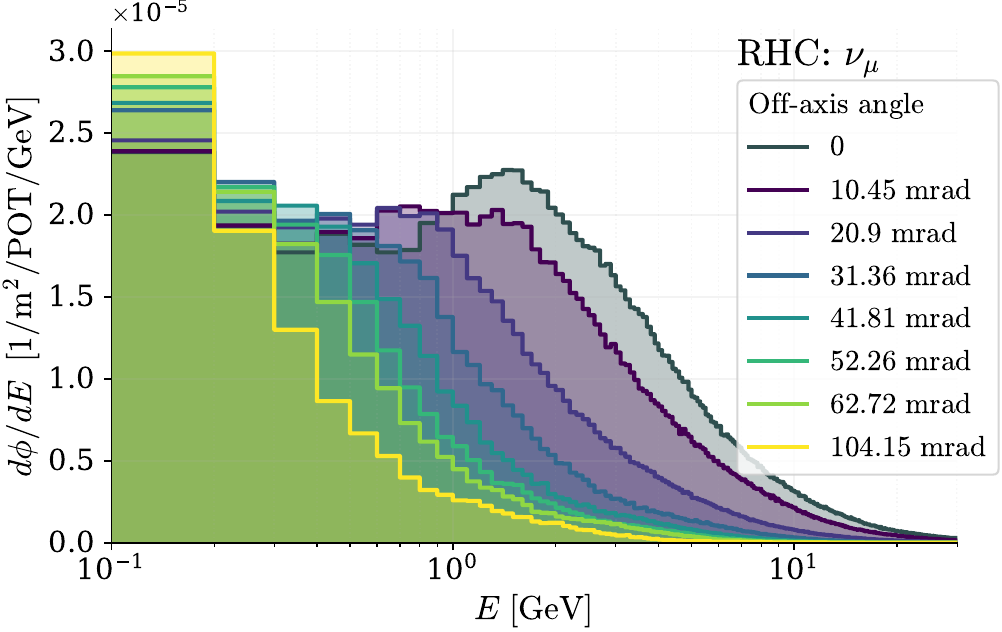}
    \includegraphics[width=0.496\textwidth]{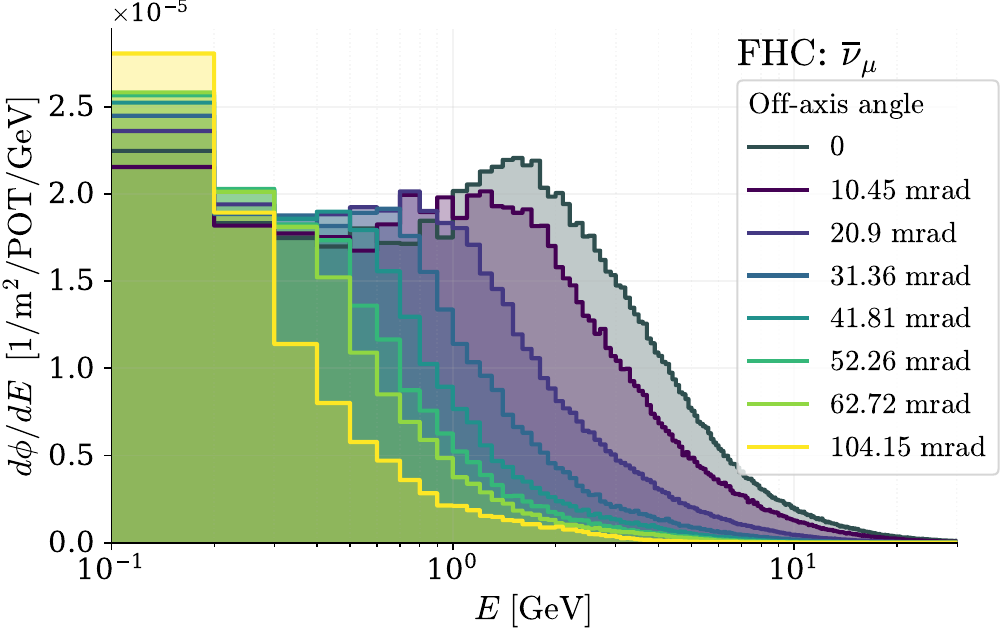}
    \includegraphics[width=0.496\textwidth]{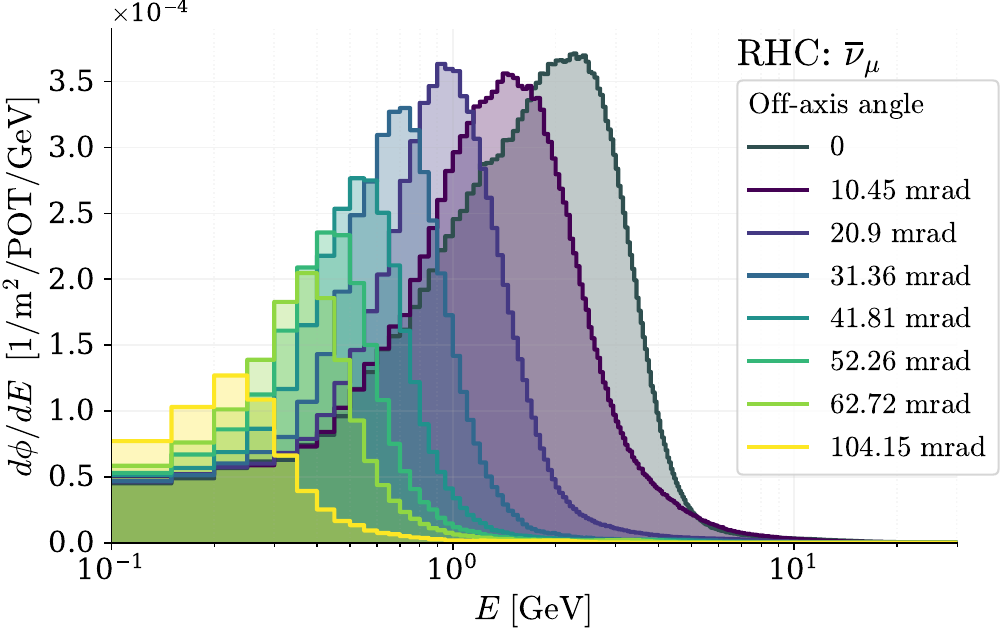}
    \includegraphics[width=0.496\textwidth]{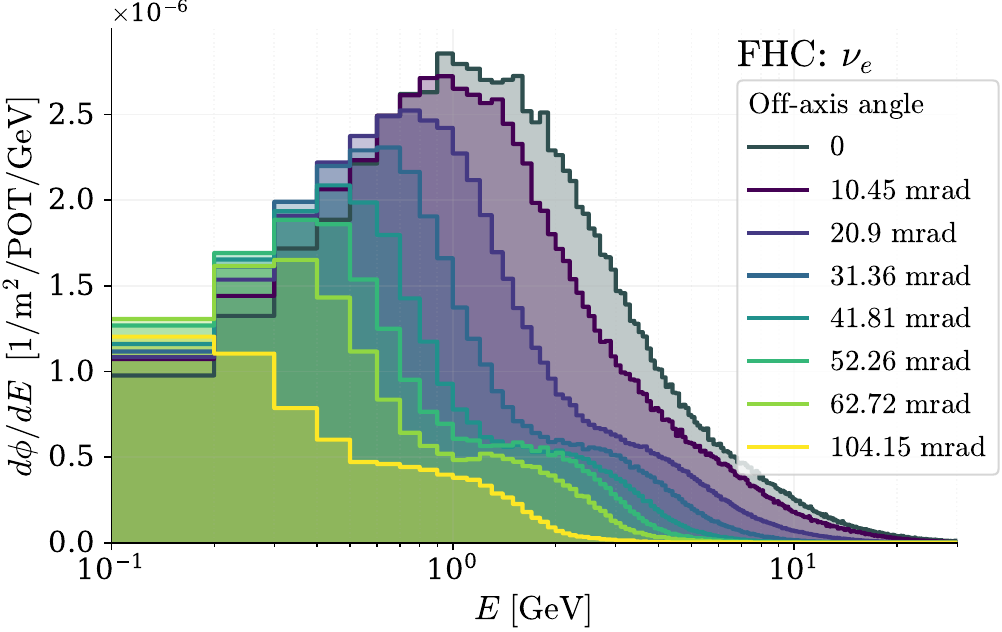}
    \includegraphics[width=0.496\textwidth]{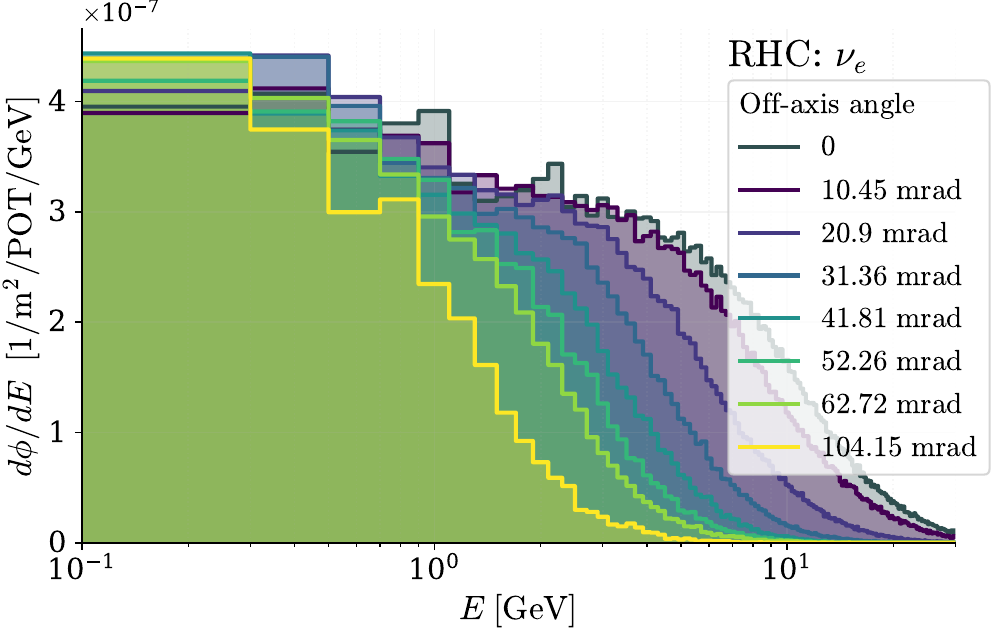}
    \includegraphics[width=0.496\textwidth]{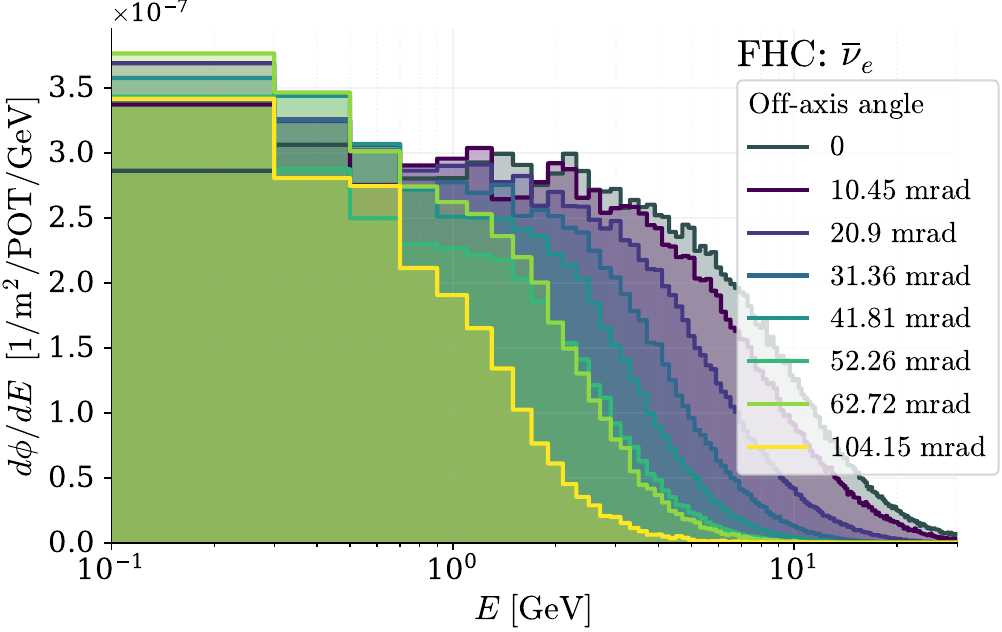}
    \includegraphics[width=0.496\textwidth]{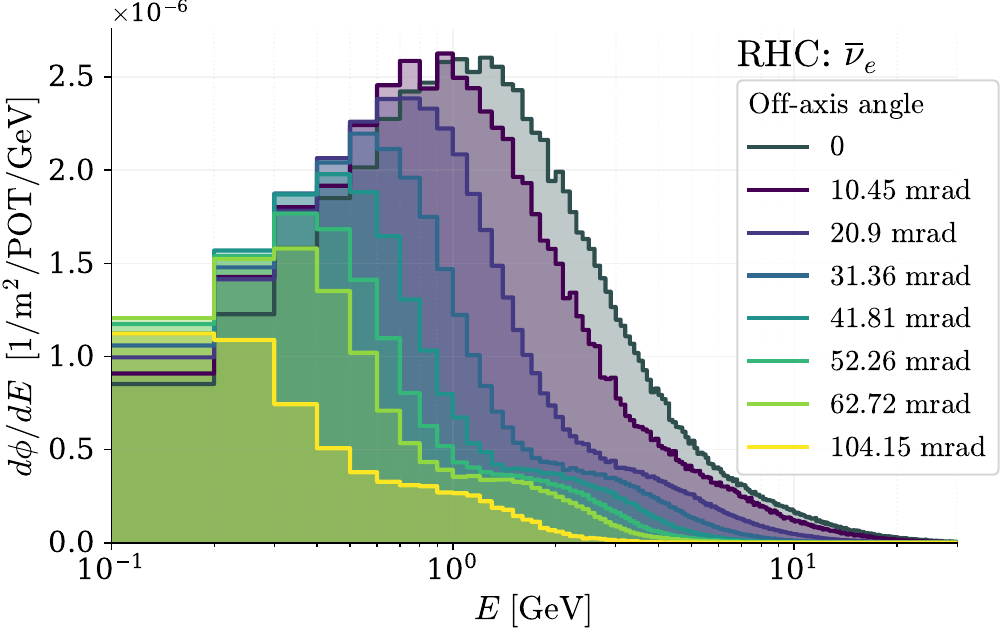}
    \caption{
    Neutrino fluxes at the DUNE Near Detector (ND-LAr) for various off-axis angles, shown as a function of neutrino energy $E$. Left panels display the fluxes for neutrino mode (FHC), while right panels show antineutrino mode (RHC). Within each column, the panels show the fluxes of $\nu_\mu$, $\overline\nu_\mu$, $\nu_e$, and $\overline{\nu}_e$, from top to bottom, respectively. The flux files are available as ancillary files to this manuscript.
    }
    \label{fig:ndlar_flux}
\end{figure}

The resulting fluxes as a function of the neutrino energy $E$ are displayed in Figure~\ref{fig:ndlar_flux} for each neutrino species $\nu_\mu$, $\overline\nu_\mu$, $\nu_e$, and $\overline{\nu}_e$, from top to bottom, respectively. Left panels show the fluxes during the neutrino mode (FHC), while the right panels show the results for the antineutrino mode (RHC). A deep charcoal shade is used to plot the fluxes in the on-axis configuration, while the results for the seven off-axis configurations are displayed scaling from dark purple (low values) to yellow (high values).

During the FHC mode, the beam is primarily composed of $\nu_\mu$ originating from the decay of the focused $\pi^+$ and $K^+$, which contribute to the peak and the tail of the spectrum, respectively. In the on-axis configuration, the spectrum peaks at 2--3~GeV, shifting towards lower energies as the off-axis angle increases. The number of $\nu_\mu$ also decreases with the off-axis angle, reducing the statistics by approximately 66\% when comparing the on-axis flux to the one at 104.15~mrad. During the neutrino mode, the beam also contains a contamination of $\overline\nu_\mu$ from the decay of the remaining $\pi^-$ that the magnetic horns are unable to deflect. The flux of $\nu_e$ arises mostly from the decay of the $\mu^+$ produced via pion decay (peak) and from the three-body decay of $K^+$ (tail). Finally, the contamination of $\overline\nu_e$ comes mostly from $K^0_L$ decays. 

In the antineutrino mode, the description of the fluxes above holds true, but with the roles of neutrinos and antineutrinos exchanged. This symmetry is broken by small differences in the magnetic fields produced by the horns, which introduce slight discrepancies in how the charged mesons are focused or deflected. In addition, it is further broken at the production level by the intrinsic asymmetry in charged meson yields in proton–target collisions: in particular $\pi^+$ and $K^+$, are produced more abundantly than their negatively charged counterparts.

In Appendix~\ref{app:tau_detection}, we also present the neutrino fluxes crossing ND-LAr with the different DUNE-PRISM off-axis configurations assumed in this section, based on the $\tau$-optimized beam mode currently under consideration by the DUNE collaboration.

\subsection{Systematic uncertainties and correlations}\label{Sec:systematics_correlations}

At this point, several sources of systematic uncertainty arise. The dominant one originates from meson production, which depends on complex hadronic interactions inside the target and beamline components while additional contributions come from imperfect knowledge of horn focusing and alignment. Consequently, the neutrino flux cannot be predicted from first principles and must instead be obtained from detailed beam Monte-Carlo simulations. These effects affect both the overall normalization of the flux and its energy dependence. 

Because the flux peaks at a few GeV, neutrino interactions occur through a mixture of quasi-elastic scattering, resonance production, and deep-inelastic processes. While high-energy scattering can be described in terms of partonic degrees of freedom, the lower-energy channels depend sensitively on nuclear dynamics. Multi-nucleon correlations and final-state interactions modify the visible final state and distort the relation between reconstructed and true neutrino energy.

The two classes of uncertainties described above are the primary reason long-baseline neutrino facilities deploy near detectors: by measuring the unoscillated flux directly, they cancel the dominant flux and cross-section systematics that would otherwise limit oscillation analyses at the far detector. For new-physics searches at the near detector, however, the same uncertainties can become the leading limitation. This is particularly true in scenarios such as non-unitarity of the PMNS matrix or the presence of light sterile neutrinos, where existing constraints from neutrino oscillation experiments limit deviations from the standard prediction to the percent level~\cite{Blennow:2025qgd}. Due to the larger normalization systematics expected at the near detector, sensitivity to these scenarios relies on differences in the spectral shape of the events. Therefore, how well the shape of the standard prediction is known will determine the ultimate reach for these scenarios.

The DUNE collaboration has provided GLoBES files~\cite{DUNE:2021cuw,Huber:2004ka,Huber:2007ji} intended for far-detector analyses within the three-neutrino oscillation framework. They consider two main event categories:
\begin{itemize}
\item \textbf{$\nu_e$-like events.} These correspond to CC $\nu_e$ interactions and are characterized by a nucleon and an electron emerging from the interaction vertex, with the electron producing an electromagnetic shower in the detector. The sample includes electron neutrinos from oscillations as well as intrinsic $\nu_e$ present in the beam. It also contains a small fraction of misidentified events from CC $\nu_\mu$, CC $\nu_\tau$, and neutral-current (NC) interactions. Misidentification may occur, for example, when a CC $\nu_\mu$ interaction produces a very short muon track, or when an NC interaction generates a $\pi^0$ whose decay photons mimic an electromagnetic shower. 
\item \textbf{$\nu_\mu$-like events.} These are characterized by a long, straight muon track emerging from the interaction vertex. In the standard 3 neutrino scenario they originate almost entirely from genuine $\nu_\mu$ and $\bar\nu_\mu$ in the beam, with smaller contributions from misidentified CC $\nu_\tau$ and NC events. Contamination from $\nu_e$ misidentification is negligible. 
\end{itemize}

On the other hand, $\nu_\tau$-like events are not included in the official files. As we will argue later, the impact of DUNE-PRISM in the $\tau$ sector is marginal, and we therefore relegate its discussion to Appendix~\ref{app:tau_detection}. The DUNE files provide detection efficiencies, misidentification probabilities, and smearing matrices relating true neutrino energy to reconstructed energy, which we use in our analysis. They also include normalization systematics. However, as discussed above, normalization uncertainties alone are insufficient for near-detector new-physics searches, where the dominant limitation arises from spectral uncertainties that are often neglected in the literature. %neglected in the standard analysis. 
The DUNE-PRISM concept qualitatively changes this situation. By sampling neutrinos at multiple off-axis angles, the experiment measures a set of narrow-band fluxes whose relative shapes are determined primarily by geometry rather than by hadron-production modeling or interaction uncertainties, since the same cross section applies across all off-axis configurations. As a result, DUNE-PRISM measurements can strongly reduce the combined impact of flux and cross-section systematics.
%%%%%%%%%%%%%%%%%%%%%
\begin{table}
\begin{center}
\begin{tabular}{cc | cc}
\hline \hline
Event sample & Contribution & $\sigma_{\rm norm}$ & $\sigma_{\rm shape}$ \\ \hline\hline
\multirow{4}{*}{$\nu_e$-like} 
& Signal              & 5\%  & -- \\
& Intrinsic cont.     & 10\% & 5\% \\
& Flavor mis-ID       & 5\%  & 5\% \\
& NC                  & 10\% & 5\% \\ \hline

\multirow{2}{*}{$\nu_\mu$-like}  
& $\nu_\mu,\bar\nu_\mu$ CC (signal) & 10\% & 5\% \\
& NC                                & 10\% & 5\% \\ \hline

\multirow{2}{*}{$\nu_\tau$-like} 
& Signal & 20\% & -- \\
& NC     & 10\% & 5\% \\ \hline \hline
\end{tabular}
\end{center}
\caption{\label{tab:sys} Assumed prior uncertainties affecting the normalization and spectral shape of the event rates in our simulations. The same uncertainties (taken as uncorrelated) are applied to both neutrino and antineutrino modes. Background contributions are divided into intrinsic beam contamination (intrinsic cont.), flavor mis-identification (flavor mis-ID) and neutral-current (NC) backgrounds. Including a shape uncertainty for the $\nu_e$-like and $\nu_\tau$-like signal has a negligible effect on the analysis.} 
\end{table}
%%%%%%%%%%%%%%%%%%%%%
To account for shape uncertainties and assess the effect that DUNE-PRISM could have on them, we adopt a similar approach to that of Ref.~\cite{Coloma:2021uhq}. We take a conservative value of shape uncertainty of $5\%$, but correlate the nuisance parameters between different off-axis modes: they remain uncorrelated bin-to-bin but fully correlated across off-axis configurations. In this way we attempt to mimic the reduction of shape systematics that DUNE-PRISM could eventually achieve. We summarize in Table~\ref{tab:sys} the systematics used in this work. It is worth noting that although we include normalization systematics with values ranging from $5\%$ to $20\%$, since they will be present, they only play a marginal role in the results\footnote{The normalization systematic uncertainty mainly affects the averaged-out sterile oscillation regime in disappearance channels ($P_{\alpha\alpha}$), which can be mapped onto non-unitarity effects. However, as already mentioned, in this case the current constraints are already stronger than the future DUNE sensitivity.}, and the truly limiting systematics for these types of new-physics searches always remain the shape uncertainties. We have explicitly checked that including larger normalization systematics does not alter our results. In Section~\ref{sec:statistical_analysis}, we will give further details on the implementation of the systematics in our statistical analysis.

\subsection{Statistical analysis}\label{sec:statistical_analysis}

To incorporate spectral (shape) systematics in the fit, we introduce a set of nuisance parameters acting independently in each reconstructed energy bin; however, we consider them to be correlated between the different off-axis modes, as discussed in \ref{Sec:systematics_correlations}.

We denote $O_{i\alpha}$ the event rates predicted in the three--neutrino framework for each $i$-bin of reconstructed energy and each off-axis mode $\alpha$, and $N_{i\alpha}$ the prediction of the tested model, depending on the model parameters ${\Theta}$. The shape nuisance parameters are denoted as $\xi_{i}^{\rm sig}$ and $\xi_{i}^{\rm bg}$ for signal and background respectively. In addition, overall normalization variations are implemented through nuisance parameters $\zeta_{c\alpha}$, assigned independently to each channel $c=s,b$ (where $s$ and and $b$ denote signal and background respectively) each off-axis mode specified by the index $\alpha$. These parameters describe a coherent rescaling of the full spectrum of a given contribution within each configuration. With these definitions of the nuisance parameters, we can write $N_{i\alpha}$ as:

\begin{equation}
\label{eq:Nch}
N_{i\alpha}(\{\Theta,\xi,\zeta\}) =
\sum_s (1+\xi_{i}^{\rm sig}+\zeta_{s\alpha})\, s_{i\alpha}(\{\Theta\})
+
\sum_b (1+\xi_{i}^{\rm bg}+\zeta_{b\alpha})\, b_{i\alpha}(\{\Theta\}) .
\end{equation}

For each sample we define a Poisson likelihood test statistic as
\begin{align}
\label{eq:chi2-full}
\chi^2(\{\Theta,\xi,\zeta\}) =
&\sum_{i,\alpha} 2 \left[
N_{i\alpha}(\{\Theta,\xi,\zeta\}) - O_{i\alpha}
+ O_{i\alpha} \ln\!\left(\frac{O_{i\alpha}}{N_{i\alpha}(\{\Theta,\xi,\zeta\})}\right)
\right]
\nonumber\\
&+
\sum_{s,\alpha} \left(\frac{\zeta_{s\alpha}}{\sigma_{{\rm norm},s}}\right)^2
+
\sum_{b,\alpha} \left(\frac{\zeta_{b\alpha}}{\sigma_{{\rm norm},b}}\right)^2
\nonumber\\
&+
\sum_{i,\alpha} \left(\frac{\xi_{i\alpha}^{\rm sig}}{\sigma_{\rm shape,sig}}\right)^2
+
\sum_{i,\alpha} \left(\frac{\xi_{i\alpha}^{\rm bg}}{\sigma_{\rm shape,bg}}\right)^2 .
\end{align}

Here $\sigma_{{\rm norm}}$ denotes the prior uncertainty associated with the normalization of each signal or background component, while $\sigma_{\rm shape}$ parametrizes the bin--to--bin spectral distortion affecting the total signal or background rate in the sample. The final test statistic is then obtain by minimizing over the nuisance parameters:
\begin{equation}
\label{eq:chi2min}
\chi^2_{\rm min}(\{\Theta\}) =
\min_{\{\xi,\zeta\}} \chi^2(\{\Theta,\xi,\zeta\}) .
\end{equation}

Intuitively, the nuisance parameters can vary during the minimization without significant penalty as long as their shifts remain small compared to the size of their priors.

\section{Results}
\label{sec:results}

This section presents our main results. We consider separately the two theoretical scenarios studied in this work: non-unitarity of the leptonic mixing matrix arising from low-scale new physics, and oscillations involving sterile neutrinos. Before presenting the results, we define the three running configurations that will be used throughout the remainder of this study:

\begin{itemize}
    \item \textbf{DUNE ND-LAr (7 years):} We assume a conservative exposure of the nominal beam corresponding to 3.5 years of running in neutrino mode and 3.5 years in antineutrino mode (DUNE Phase I).

    \item \textbf{DUNE ND-LAr $+$ PRISM (7+7 years):} In addition to the nominal DUNE running, we assume an extra 7 years of PRISM operation (during a possible DUNE Phase II). This additional exposure is distributed equally among the seven off-axis positions and between neutrino and antineutrino modes.

    \item \textbf{DUNE ND-LAr $+$ PRISM optimized (7+7 years)\footnote{This represents a limiting scenario, in which the statistics relevant for the standard oscillation analysis are assumed to saturate after a relatively short running period, allowing the remaining time to be devoted to BSM searches such as those considered in this work. }:} As in the previous configuration, we consider the nominal DUNE running plus an additional 7 years of PRISM data. However, in this case the extra exposure is allocated to the first three off-axis positions, which are the most relevant for the present analysis.
\end{itemize}

\subsection{Non-unitarity}

As discussed in Section~\ref{Sec:systematics_correlations}, in a more general context, current constraints on new physics in neutrino oscillations allow only for small deviations from the three--neutrino prediction and in order to surpass existing bounds we require observables that produce signatures distinguishable from the standard prediction.

In the case of non-unitarity, as shown in Section~\ref{sec:theory}, the diagonal parameters $\alpha_{\beta\beta}$ only modify disappearance probabilities at the near detectors as $P_{\beta\beta} = 1 - 4\alpha_{\beta\beta}$\footnote{Note that in the standard non-unitarity scenario in which it is induced by a heavy sector, the dependence on $\alpha_{\beta\beta}$ vanishes due to normalization effects~\cite{Blennow:2016jkn,Blennow:2025qgd}.}. This corresponds simply to an overall normalization change in the $\nu_e$ and $\nu_\mu$ fluxes. The sensitivity is therefore limited by normalization systematics, which are larger than the present experimental constraints. For this reason we do not consider disappearance channels in our non-unitarity analysis. By contrast, the off-diagonal parameters $\alpha_{\beta\gamma}$ modify appearance probabilities $P_{\beta\gamma} = 2|\alpha_{\beta\gamma}|^2$. In this case the observable effect is a distortion of the event spectrum, originating from the different spectra of the parent mesons. For example for $\alpha_{\mu e}$, the $\nu_\mu$ flux is dominated by two-body pion decays, whereas the $\nu_e$ flux mainly arises from three-body kaon and muons decays, since the pion decay to electrons is helicity suppressed. Therefore, a transition $P_{\mu e}$ produces a $\nu_e$ component carrying the spectral shape characteristic of two-body pion decay, allowing it to be distinguished from the standard signal.

\begin{figure}[t!]
    \centering
    % ---------- Top panel ----------
    \includegraphics[width=0.78\linewidth]{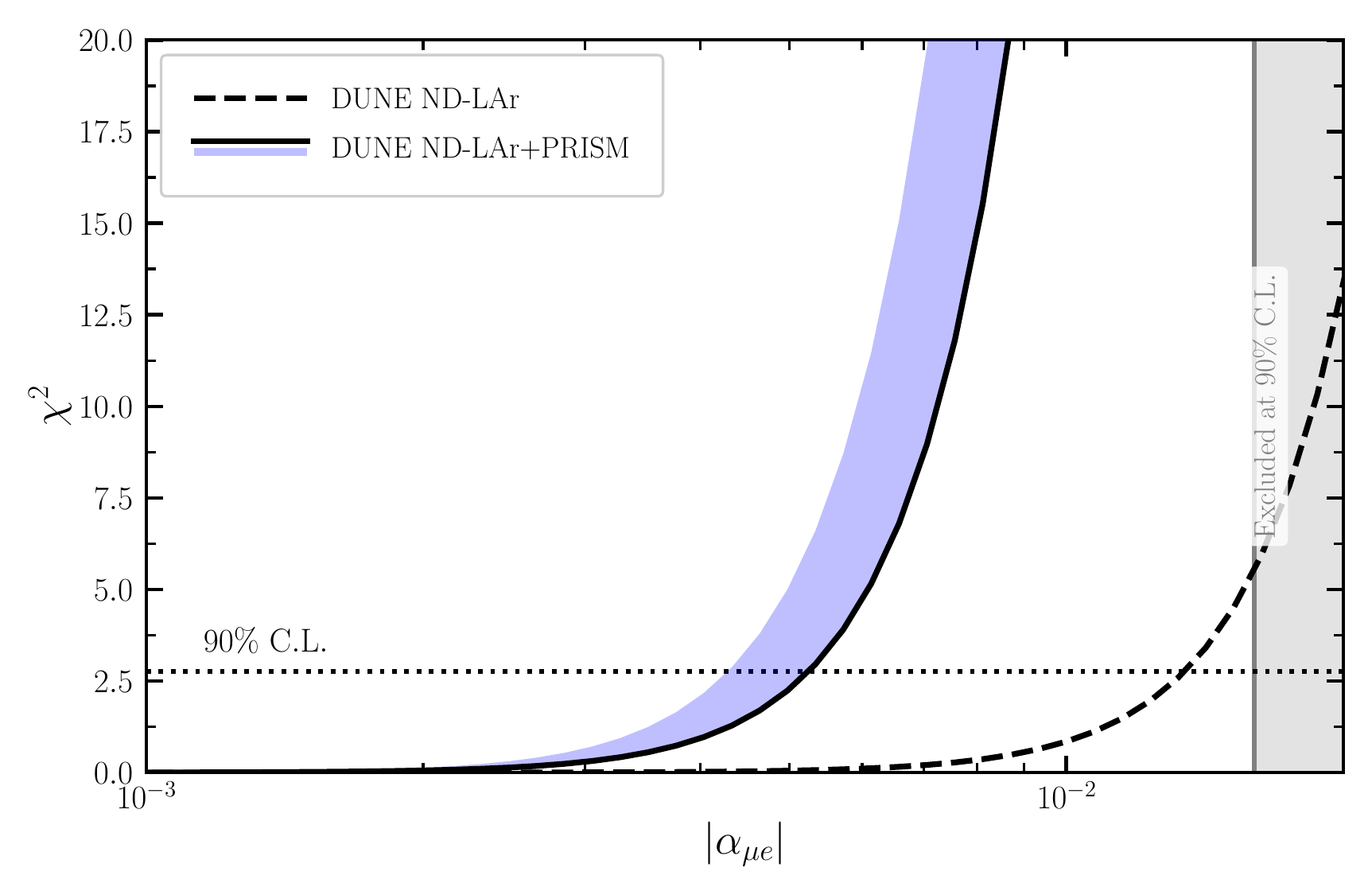}

    \caption{
Sensitivity to the off-diagonal non-unitarity parameters $\vert \alpha_{\mu e}\vert$, assuming a $5\%$ shape uncertainty. The dashed curve corresponds to the configuration without DUNE-PRISM, while the solid curve shows the sensitivity with DUNE-PRISM. The shaded blue region indicates the increase in sensitivity obtained with the DUNE PRISM optimized mode (see text for details). The horizontal line indicates the $90\%$ C.L. threshold, while the grey-shaded region denotes the parameter space excluded at $90\%$ C.L. by current constraints.
    }
    \label{fig:alpha_mue}
\end{figure}

Our results for $\vert\alpha_{\mu e}\vert$ are presented in Figure~\ref{fig:alpha_mue}, which shows $\chi^2$ as a function of $\vert\alpha_{\mu e}\vert$. The solid curve corresponds to DUNE near detector ND-LAr alone, assuming a conservative $5\%$ spectral uncertainty, yielding a sensitivity close to the present bound. The dashed curve shows that the sensitivity improves significantly once DUNE-PRISM is included, roughly by a factor of two relative to DUNE alone, comparable to the improvement obtained in Ref.~\cite{Coloma:2021uhq} for very small shape uncertainties. The final level of spectral uncertainty achievable at the DUNE near detector is still unknown. However, these results indicate that even under conservative assumptions, a relatively short off-axis run with DUNE-PRISM can substantially alleviate the limitation imposed by shape systematics.

\subsection{Sterile neutrinos 3+1}

In this section we present the future sensitivity of the DUNE near detector ND-LAr and DUNE-PRISM to sterile neutrino oscillations. For simplicity, and to facilitate comparison with existing results, we adopt the 3+1 scenario.

As discussed above, distinguishing new physics from the standard three-neutrino prediction requires more than an overall normalization shift; an energy-dependent distortion is essential. In the case of non-unitarity, only appearance channels can be probed beyond normalization effects. In contrast for sterile neutrinos with mass splittings in the range $10^{-1}\,\text{eV}^2 < \Delta m^2_{41} < 100\,\text{eV}^2$, the presence of the sterile state induces an intrinsic energy dependence in the oscillation probabilities, providing the necessary handle to disentangle it from the standard scenario in both appearance and disappearance channels.

\begin{figure}[t!]
    \centering

    % ---------- Top panel ----------
    \includegraphics[width=0.78\linewidth]{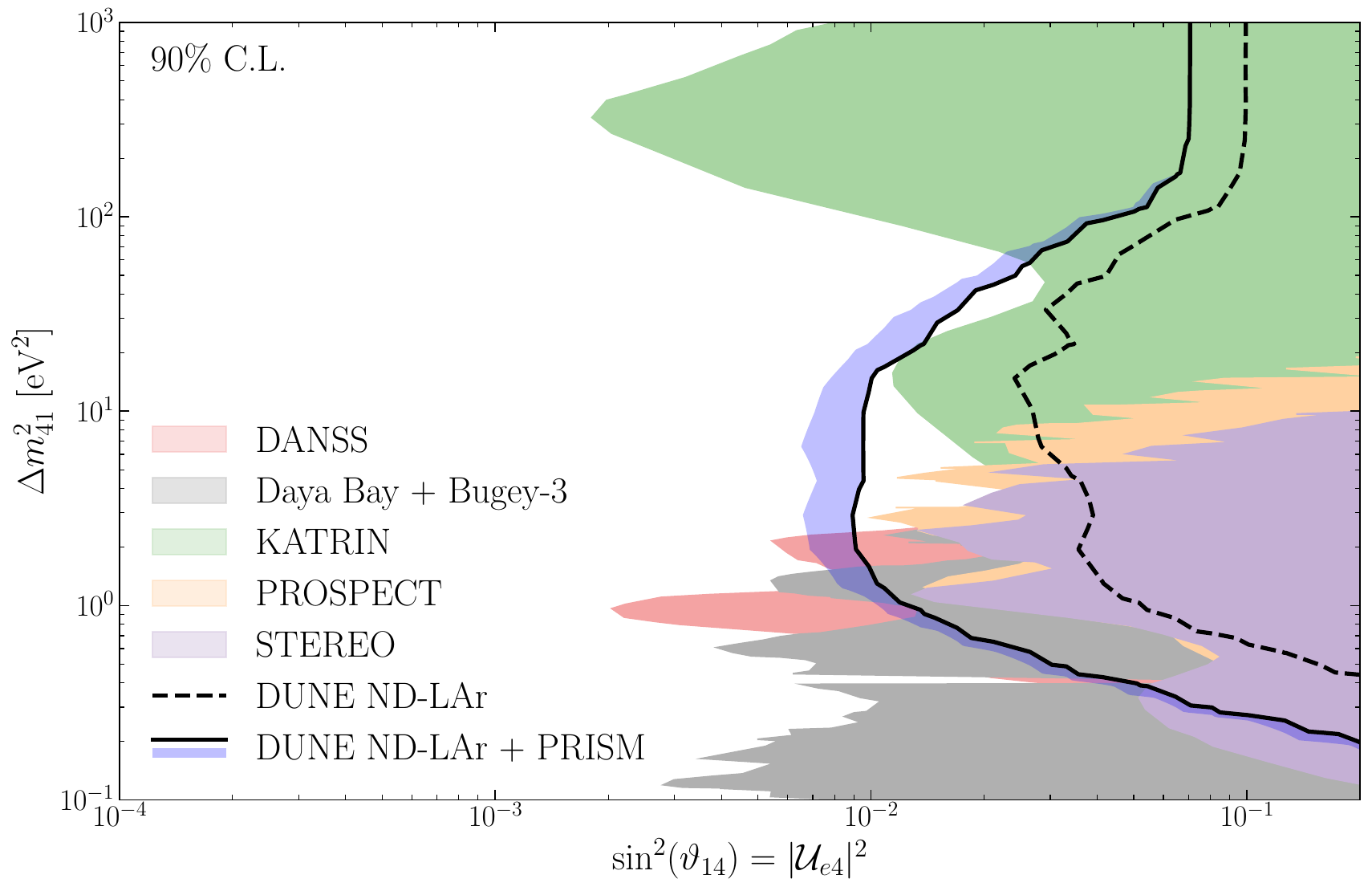}

    \vspace{0.4cm}

    % ---------- Bottom panel ----------
    \includegraphics[width=0.78\linewidth]{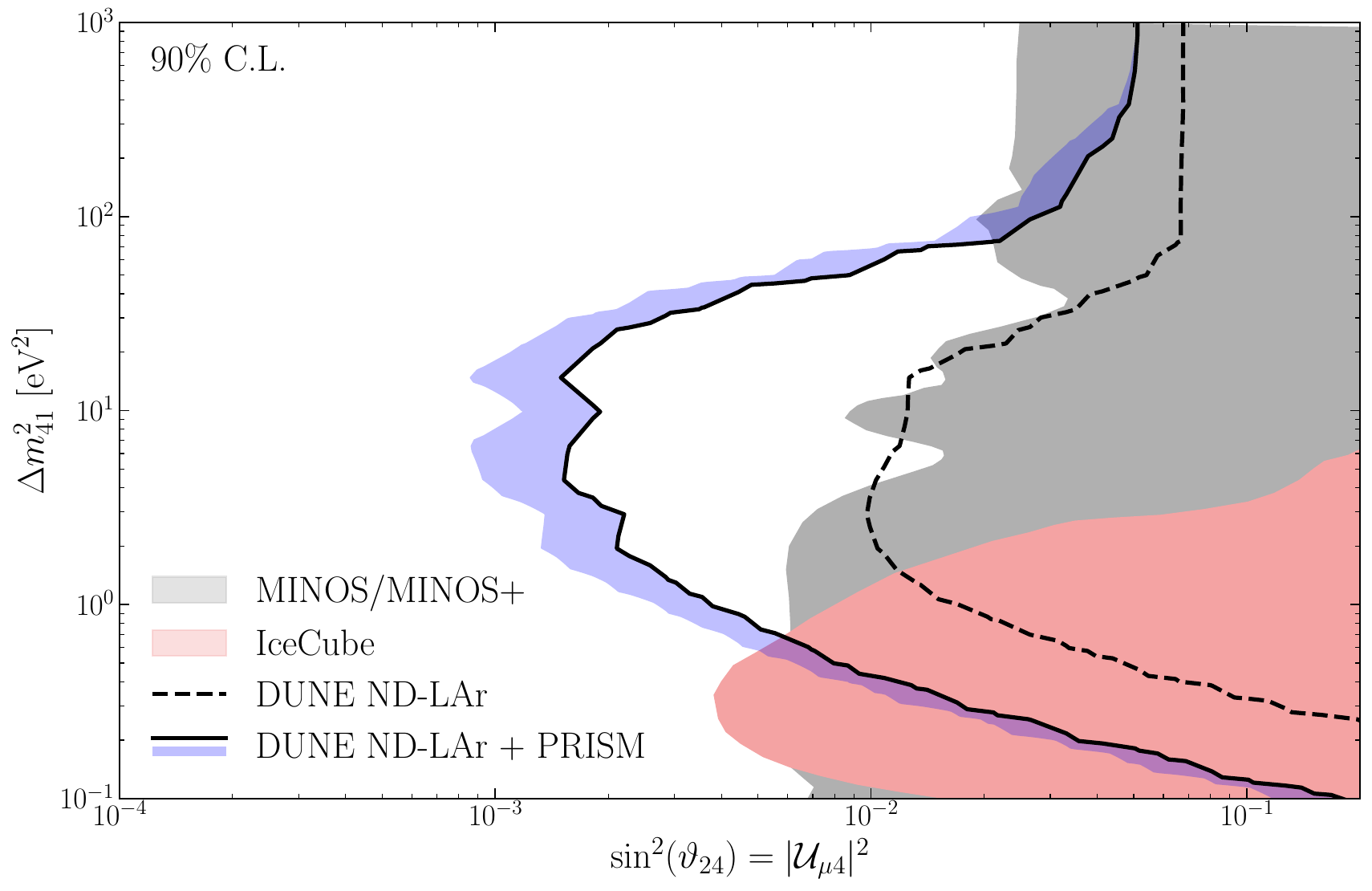}

    \caption{
   Expected sensitivity to sterile-neutrino disappearance channels assuming a $5\%$ shape uncertainty. The upper panel shows $P_{e e}$ and the lower panel shows $P_{\mu \mu}$. The shaded regions indicate current constraints from other experiments, reported at $90\%$ C.L. (IceCube, MINOS, Daya Bay and Bugey3) and $95\%$ C.L. (DANSS, PROSPECT, STEREO, KATRIN)~\cite{TheIceCube:2016oqi,Adamson:2020jvo,Machikhiliyan:2022muh,PROSPECT:2024gps,STEREO:2022nzk,KATRIN:2025lph}. The solid (dashed) line shows $90\%$ CL sensitivity with (without) DUNE-PRISM. %The solid curve shows the DUNE \jb{ND-Lar} alone sensitivity, while the dashed curve shows the sensitivity with DUNE-PRISM. In all cases, the region to the right of the curves is excluded at $90\%$ CL (2 d.o.f.). 
   The shaded blue region indicates the increase in sensitivity obtained with the DUNE PRISM optimized mode (see text for details).
    }
    \label{fig:sterile_diss}
\end{figure}

Before presenting the results, we introduce a convenient notation widely used in the literature, which also facilitates comparison with other studies. In the 3+1 scenario, the full mixing matrix can be parametrized as
$\mathcal{U} = R_{34}\, S_{24}\, S_{14}\, R_{23}\, S_{13}\, R_{12},$ where $R_{ij}$ denotes a real rotation in the $ij$ plane with mixing angle $\theta_{ij}$, while $S_{ij}$ are complex rotations that include the corresponding CP phase $\delta_{ij}$. The additional mixing angles, which control the admixture of the mostly sterile state with the active neutrinos, can be straightforwardly mapped onto the triangular parametrization, and can also be expressed in terms of the mixing matrix elements as:
\begin{equation}
\begin{aligned}
|\mathcal{U}_{e4}|^2 &= 2\alpha_{ee}=s_{14}^2,\\
|\mathcal{U}_{\mu 4}|^2 &= 2\alpha_{\mu\mu}=s_{24}^2+\mathcal{O}(s_{24}^2 s_{14}^2),\\
|\mathcal{U}_{\tau 4}|^2 &= 2\alpha_{\tau\tau}=s_{34}^2+\mathcal{O}(s_{24}^2 s_{34}^2)+\mathcal{O}(s_{34}^2 s_{14}^2).
\end{aligned}
\end{equation}

Figure~\ref{fig:sterile_diss} shows the projected $90\,\%$ C.L.\ sensitivity (2 d.o.f.) to sterile-neutrino disappearance in the $P_{ee}$ (upper panel) and $P_{\mu\mu}$ (lower panel) channels, given by their contributions to the $\nu_e$-like and $\nu_\mu$-like samples, respectively. Solid lines correspond to the standard on-axis DUNE ND-LAr configuration assuming a conservative $5\,\%$ uncorrelated bin-to-bin shape uncertainty. Dashed lines display the sensitivity achieved with the full DUNE-PRISM data set, where the same shape nuisance parameters are fully correlated across the seven off-axis positions. Current constraints from MINOS/MINOS+~\cite{MINOS:2016viw}, Daya Bay+Bugey-3~\cite{Adamson:2020jvo}, PROSPECT~\cite{PROSPECT:2024gps}, DANSS~\cite{Machikhiliyan:2022muh}, STEREO~\cite{STEREO:2022nzk}, KATRIN~\cite{KATRIN:2025lph} and IceCube ~\cite{TheIceCube:2016oqi} are shown for comparison. The improvement provided by DUNE-PRISM is most significant for $0.1 \lesssim \Delta m_{41}^2 \lesssim 100\,\mathrm{eV}^2$, where the sterile oscillation develops over the near-detector baseline. For smaller mass splittings the phase remains too small to produce a visible distortion (analogous to standard three-neutrino oscillations at short baselines), while for larger values the rapid oscillations average out, recovering the non-unitarity limit discussed earlier. However, the resulting sensitivity in this limit remains weaker than existing bounds.

\begin{figure}[t!]
    \centering

    % ---------- Bottom panel ----------
    \includegraphics[width=0.78\linewidth]{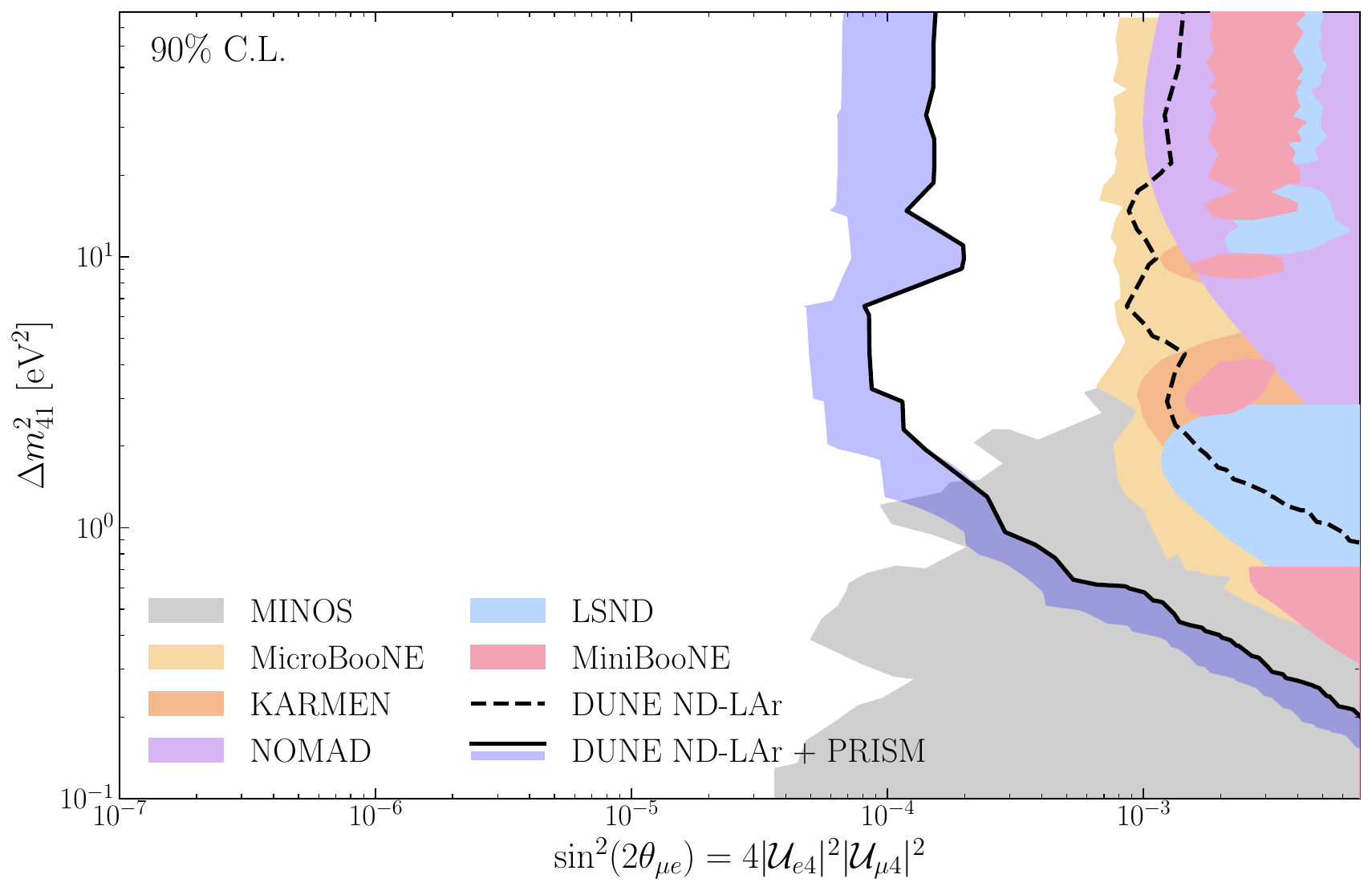}

    \caption{Expected sensitivity to sterile neutrinos from the combination of appearance and disappearance channels $P_{\mu\mu} + P_{ee} + P_{\mu e}$, assuming a $5\%$ shape uncertainty. The shaded pink, orange, yellow, and gray areas show current constraints at $90\%$ C.L. from other experiments (MINOS+, KARMEN and NOMAD)~\cite{Armbruster:2002mp,Astier:2003gs,Adamson:2020jvo,Astier:2001yj,Tsenov:2009jca}, and at $95\%$ C.L. from MicroBooNE~\cite{MicroBooNE:2025nll}, while the shaded purple regions are favored at $99\%$ C.L. by the LSND~\cite{Aguilar:2001ty} and MiniBooNE~\cite{Aguilar-Arevalo:2020nvw} anomalies. The solid (dashed) line shows $90\%$ CL sensitivity with (without) DUNE-PRISM. The shaded blue region indicates the increase in sensitivity obtained with the DUNE PRISM optimized mode (see text for details).
    }
    \label{fig:sterile_app}
\end{figure}

The combined analysis including both disappearance and appearance channels ($P_{\mu\mu} + P_{ee} + P_{\mu e}$) is shown in Figure~\ref{fig:sterile_app}. The $\nu_e$-like events provide simultaneous sensitivity to $P_{ee}$ and $P_{\mu e}$, while $\nu_\mu$-like events contribute to $P_{\mu\mu}$. Results are presented in terms of the mixing combination directly probed by the appearance channel $P_{\mu e}$, namely $4|\mathcal{U}_{e4}|^2|\mathcal{U}_{\mu4}|^2$, after marginalizing over $|\mathcal{U}_{e4}|$ and $|\mathcal{U}_{\mu4}|$. The dashed curves once again highlight the significant improvement achieved with the inclusion of DUNE-PRISM. The shaded regions denote parameter space excluded by NOMAD~\cite{Astier:2001yj}, KARMEN~\cite{Armbruster:2002mp}, MINOS/MINOS+~\cite{MINOS:2016viw}, MicroBooNE~\cite{MicroBooNE:2025nll} and IceCube~\cite{TheIceCube:2016oqi}, while the orange and yellow regions correspond to the areas favored at $99\,\%$ C.L.\ by the LSND and MiniBooNE anomalies. The impact of including DUNE-PRISM is very significant with an improvement on the sensitivity to the mixing squared of one order of magnitude stronger than standard set up with DUNE ND-LAr alone.

\section{Summary and conclusions}
\label{sec:summary}

Precision long-baseline facilities such as DUNE will provide an unprecedented control of neutrino oscillations, but their ultimate physics reach, in particular for new-physics searches at the near detector (ND), is tightly linked to how well neutrino fluxes and neutrino-nucleus interactions can be modeled. In this work we have explored the impact of DUNE-PRISM in searches of physics beyond the three-neutrino paradigm, focusing on two representative scenarios: low-scale non-unitarity of the effective $3\times3$ mixing matrix and light sterile-neutrino oscillations in the minimal $3+1$ scenario.

A recurring theme across these scenarios is that, once current neutrino oscillation experiments constrain new effects to the percent level or below, the large normalization systematics affecting both neutrino fluxes and cross sections imply that new-physics signals producing only a modification of the overall rate can not be distinguished from the background. %an overall rate change are very limited. 
As a result, the identification of new physics requires an additional energy dependence to distinguish it from the standard three-neutrino prediction. In this context, normalization systematics at the ND become subdominant, and the sensitivity is instead primarily limited by uncertainties in the spectral shape. We have implemented these effects through bin-to-bin nuisance parameters and find that their inclusion significantly degrades the sensitivity. To alleviate this limitation, we model the expected impact of the PRISM (Precision Reaction Independent Spectrum Measurement) technique, which exploits measurements at multiple off-axis angles. This provides a data-driven handle on the neutrino energy spectrum, effectively reducing spectral uncertainties by correlating shape distortions across different off-axis configurations. In order to perform our study we have computed the neutrino fluxes arriving to the detector for the different off-axis positions considered. 
We have made all neutrino fluxes generated for this study publicly available as ancillary files accompanying this work.

For low-scale non-unitarity, disappearance channels do not provide additional energy dependence, and the sensitivity is therefore limited by normalization systematics. We thus focus on the appearance channel $P_{\mu e}$, which at very short baselines is controlled by $\alpha_{\mu e}$. In this case, the $\nu_\mu\to\nu_e$ transition injects a pion-decay-like spectral shape into the $\nu_e$ sample, which can be discriminated from the intrinsic $\nu_e$ component. Under a conservative $5\%$ shape uncertainty, we find that DUNE-PRISM can substantially improve the sensitivity to $\alpha_{\mu e}$, reaching roughly a factor-of-two gain relative to the on-axis-only configuration and restoring a performance comparable to that obtained with much smaller spectral uncertainties.

For sterile neutrinos, the new oscillation induced by the sterile state provides an additional energy dependence that can be exploited in both disappearance and appearance channels for $10^{-1}\mathrm{eV}^2 \lesssim \Delta m_{41}^2 \lesssim 100 \mathrm{eV}^2$, allowing it to be distinguished from the standard three family prediction. In this scenario, we find that DUNE-PRISM yields the largest improvements, enhancing the projected reach in $P_{ee}$ and $P_{\mu\mu}$ and, in the combined fit, significantly improving the sensitivity in the %plane 
relevant combination of mixing parameters for appearance $4|\mathcal{U}_{e4}|^2|\mathcal{U}_{\mu4}|^2$ by approximately one order of magnitude. The gain relative to the on-axis-only configuration is again comparable to that obtained under the assumption of small shape uncertainties.

Finally, we have also examined the $\tau$ sector, both for non-unitarity and sterile-neutrino induced $\nu_\mu\to\nu_\tau$ transitions, focusing on hadronic $\tau$ decays. In this case, the benefit of DUNE-PRISM is limited since most off-axis flux configurations populate energies below the $\tau$ production threshold, so the off-axis data do not provide the same leverage on the relevant signal region. Consequently, improvements remain marginal compared to the $e$ and $\mu$ sectors, and we have relegated this discussion to Appendix~\ref{app:tau_detection}, where we also consider the possibility of using a $\tau$-optimized beam.

In summary, our results highlight that spectral systematics are the dominant obstacle for percent-level new-physics searches in oscillations at the DUNE near detector, and that DUNE-PRISM offers a robust, experimentally grounded strategy to mitigate these limitations. Even under conservative assumptions on shape uncertainties, the inclusion of a PRISM program can markedly sharpen DUNE's sensitivity to non-unitarity and sterile-neutrino scenarios in the electron and muon sectors, improving the prospects for testing short-baseline new physics with next-generation long-baseline infrastructure.

%\medskip

\section*{Acknowledgments.} 

We would like to thank Justo Martin-Albo for useful discussions. We acknowledge financial support from European Union’s Horizon 2020 research and innovation program under the Marie Skłodowska-Curie grants HORIZON-MSCA-2021-SE-01/101086085-ASYMMETRY and H2020-MSCA-ITN-2019/860881-HIDDeN. \\JLP and JHG also acknowledge partial support by the Spanish Severo Ochoa Excellence grant CEX2023-001292-S (AEI/10.13039/501100011033), from the Spanish Research Agency (Agencia Estatal de Investigaci\'on) through grants PID2023-148162NB-C21 and PID2022-137268NA-C55, and CNS2022-136013 funded by MICIU/AEI/10.13039/501100011033 and by “European Union NextGenerationEU/PRTR''. Finally, JHG warmly thanks the hospitality of the TH division during his stay at CERN; where this project has been completed. 

\appendix

\section{$\nu_\tau$ detection} \label{app:tau_detection}

In this appendix we present our results for non-unitarity and sterile neutrinos in the context of $\nu_{\tau}$ detection. As we have already argued, in this case DUNE-PRISM does not effectively 
bring a significant reduction of the relevant systematic uncertainties in these scenarios and, therefore, alternative strategies must be considered. Although the results in the $\tau$ sector are less favorable, we consider it important to discuss them for completeness and in order to clarify why this approach is ineffective in this scenario. 

\subsection{Simulation details for $\nu_\tau$ detection}

The $\tau$ sector poses two main challenges. First, due to the large mass of the $\tau$ lepton, its production requires neutrino energies above threshold, $E_{\text{th}} \gtrsim 3.5~\mathrm{GeV}$. Since the DUNE neutrino flux peaks around $2$--$3~\mathrm{GeV}$, only a small fraction of the neutrino spectrum is kinematically able to produce $\tau$ leptons. Second, the very short lifetime of the $\tau$ makes $\nu_\tau$ interactions intrinsically difficult to identify experimentally.

Regarding the first limitation, the DUNE collaboration has explored a $\tau$-optimized beam configuration~\cite{ntuples}, in which the focusing horns are tuned to shift the neutrino flux peak towards $3$--$4~\mathrm{GeV}$. This enhances the fraction of neutrinos above the $\tau$ production threshold and significantly increases the expected $\nu_\tau$ event rate.

Concerning detection, $\nu_\tau$ events are not included in the standard oscillation analyses of Ref.~\cite{DUNE:2021cuw}, and are therefore absent from the official DUNE GLoBES configuration files.\footnote{We use the SM CC $\nu_\tau$ cross section from the auxiliary material of Ref.~\cite{Alion:2016uaj}, generated with GENIE v2.8.4~\cite{Andreopoulos:2015wxa,Tena-Vidal:2021rpu}.} Nevertheless, in the presence of new physics scenarios such as non-unitary mixing or sterile neutrinos, the transition probability $P_{\mu\tau}$ can become sizable, rendering $\nu_\tau$ appearance a relevant probe. The identification of $\nu_\tau$ interactions in LArTPC detectors is particularly challenging. As we have already mentioned, at neutrino energies of a few GeV, the $\tau$ lepton produced in charged-current interactions travels only a few millimetres before decaying, preventing direct track reconstruction. We therefore model these events following the approach of Refs.~\cite{Coloma:2021uhq,DeGouvea:2019kea,Kopp:2025ffx}. Since the $\tau$ itself cannot be reconstructed, the signal must be inferred from its decay products. We focus on hadronic decay modes, which account for approximately $65\%$ of $\tau$ decays, %while 
since leptonic channels ($\tau^{-} \to \ell^{-} \bar{\nu}_{\ell} \nu_\tau$), each with branching ratio $\sim 17\%$~\cite{Zyla:2020zbs}, are expected to be difficult to distinguish from standard CC $\nu_e$ and $\nu_\mu$ interactions. We assume a conservative signal efficiency of $30\%$. The reconstructed neutrino energy is degraded by the missing energy carried away by neutrinos in the $\tau$ decay. We model this effect by assigning to a true neutrino energy $E_\nu^{\text{true}}$ a Gaussian reconstructed distribution centered at $0.45\,E_\nu^{\text{true}}$ with a width of $0.25\,E_\nu^{\text{true}}$~\cite{DeGouvea:2019kea}. The dominant background arises from neutral-current interactions producing hadronic activity with missing energy. We assume a constant misidentification probability of $0.5\%$~\cite{DeGouvea:2019kea}, and account for energy smearing using the NC migration matrices provided in Ref.~\cite{DUNE:2021cuw}.

\subsection{Neutrino fluxes at DUNE-PRISM using the $\tau$-optimized beam}
%%%%%%%%%%%%%%%%%%%%%%%%%%%%%%%%%%%%%%%%

\begin{figure}[p]
    \centering
    \includegraphics[width=0.496\textwidth]{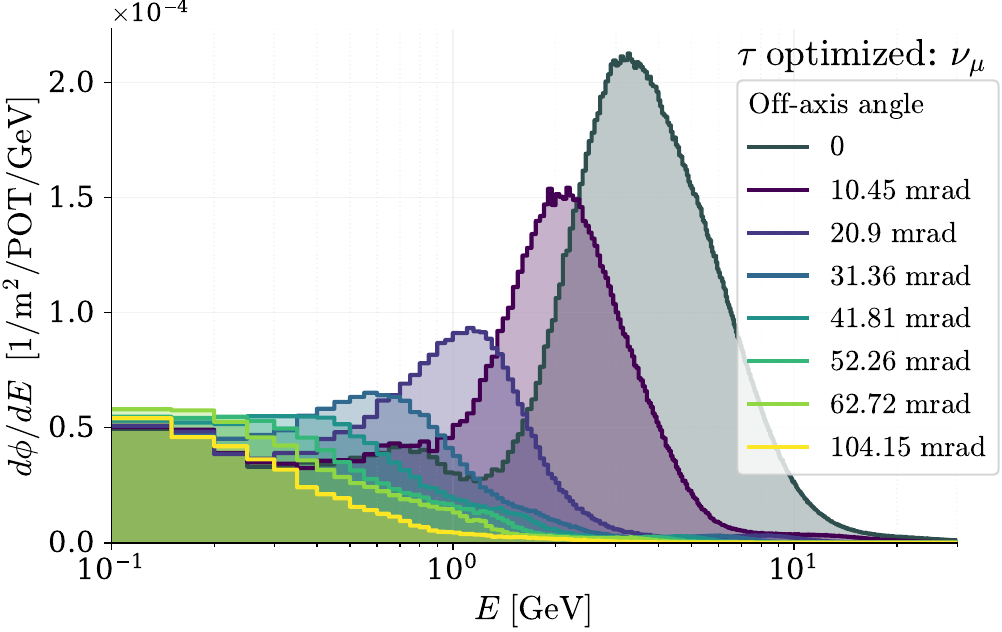}
    \includegraphics[width=0.496\textwidth]{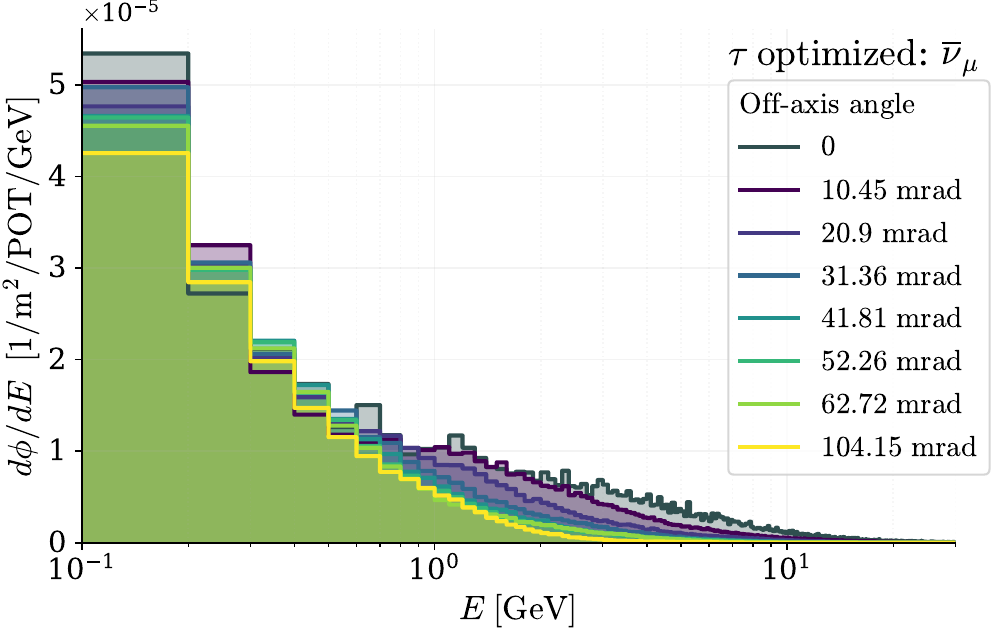}
    \includegraphics[width=0.496\textwidth]{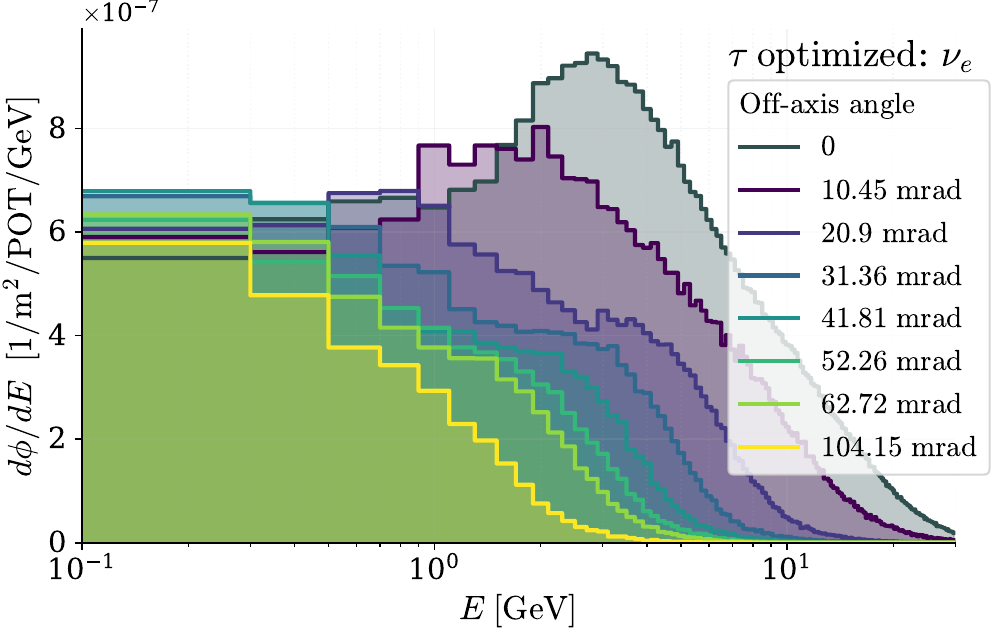}
    \includegraphics[width=0.496\textwidth]{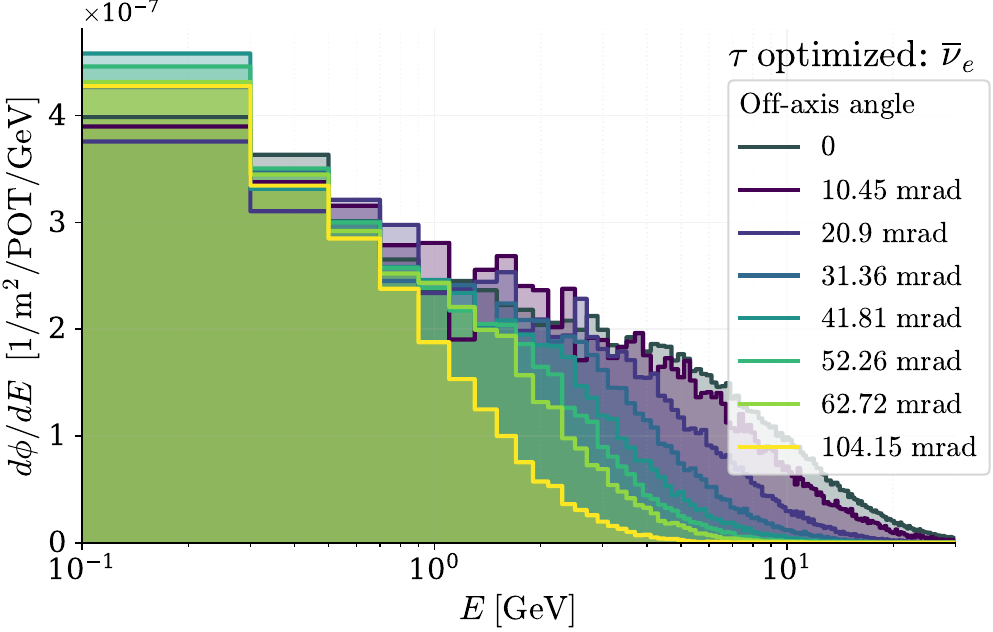}
    \caption{
    Neutrino fluxes at the DUNE Near Detector (ND-LAr) for various off-axis angles, shown as a function of neutrino energy $E$ using the $\tau$-optimized beam. The panels show the fluxes of $\nu_\mu$, $\overline\nu_\mu$, $\nu_e$, and $\overline{\nu}_e$, respectively. The fluxes of $\nu_\tau$ and $\overline\nu_\tau$ are subleading since they only come from decays of the few $D$/$D_s$ produced at the target and absorber.
    Flux files are available as ancillary files to this manuscript.}
    \label{fig:ndlar_tauoptimized_flux}
\end{figure}

The DUNE collaboration is also considering a separate run using a target and magnetic horn tuned for a beam with a higher-energy neutrino flux~\cite{DUNE:2020ypp}. This configuration would enhance $\nu_\tau$ detection in the FD and increase sensitivity for exploring BSM physics. To simulate the light neutrino fluxes crossing ND-LAr at different off-axis configurations, we started with the corresponding G4LBNF TTree tarballs\footnote{The $\tau$-optimized version of G4LBNF corresponds to the January 2021 simulations, featuring a 1~m NuMI fin-style target and NuMI-style horns.} and followed the same procedure explained in Section~\ref{sec:neutrino_flux}. Figure~\ref{fig:ndlar_tauoptimized_flux} displays our resulting fluxes for the off-axis distances introduced in Table~\ref{tab:off-axis}. The most characteristic change relative to the fluxes shown in Figure~\ref{fig:ndlar_flux} is the spectral shift toward higher energies; in particular, the $\nu_\mu$ flux peak shifts toward 3–4~GeV. 

\subsection{Results fo $\nu_\tau$ detection}
%%%%%%%%%%%%%%%%%%%%%%%%%%%%%%%%%%%%%%%%

For the $e$ and $\mu$ sectors, the multiple off-axis measurements allow for several correlated observations in which the only differences arise from kinematics and from the systematics associated with horn modeling, which are substantially smaller than hadronic or neutrino cross-section uncertainties. However, moving off-axis also implies that only lower-energy neutrinos can reach the detector: as the energy of the parent meson increases, the decay cone becomes more collimated. As a consequence, the off-axis flux peaks at lower energies—the larger the off-axis angle, the lower the energy. This implies that, for $\tau$ detection, most off-axis events fall below the production threshold and therefore do not contribute to reducing spectral systematics in the $\nu_\tau$ channel; only a marginal improvement can be expected.

\begin{figure}[t]
    \centering

 \includegraphics[width=0.78\linewidth]{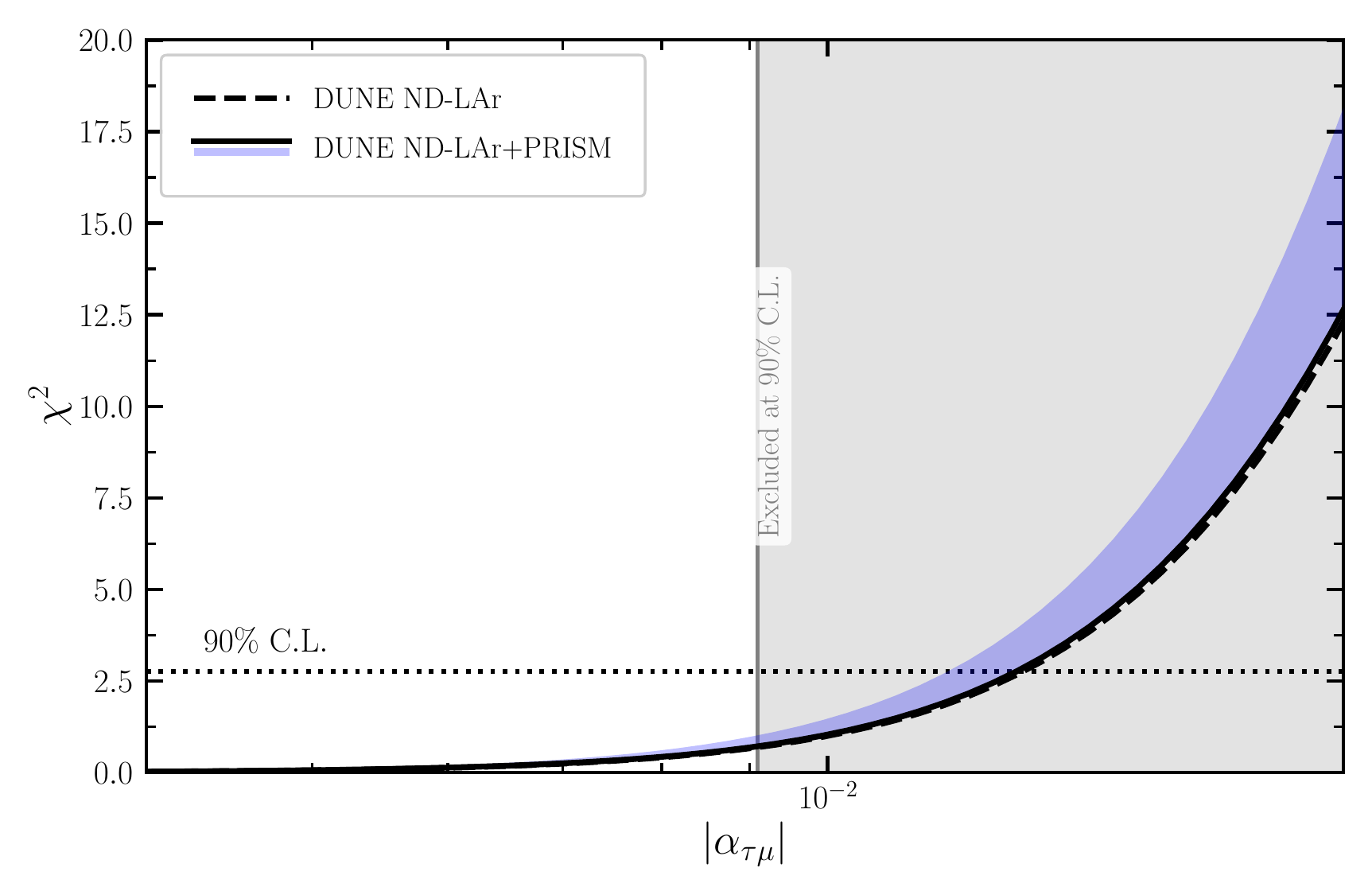}

    \caption{
Sensitivity to the off-diagonal non-unitarity parameters $\vert\alpha_{\tau\mu}\vert$, assuming a $5\%$ shape uncertainty. The dashed curve corresponds to the configuration without DUNE-PRISM, while the solid curve shows the sensitivity with DUNE-PRISM. The shaded blue region indicates the increase in sensitivity obtained with an additional run using a $\tau$-optimized beam. The horizontal line indicates the $90\%$ C.L. threshold, while the grey-shaded region denotes the parameter space excluded at $90\%$ C.L. by current constraints.}
\label{fig:alpha_mutau}
\end{figure}
In Figure~\ref{fig:alpha_mutau}, we present the analogous results to those shown in Figure~\ref{fig:alpha_mue}, now for $\alpha_{\tau\mu}$. The solid curve corresponds to DUNE ND-LAr alone, assuming a %conservative 
$5\%$ spectral uncertainty, and yields a sensitivity worse than the current bound. This sensitivity is not significantly improved by the inclusion of DUNE-PRISM (dashed curve), for the reasons discussed above. The shaded blue region indicates the additional gain in sensitivity obtained when including a dedicated run with the $\tau$-optimized beam configuration. 

\begin{figure}[t]
    \centering

    % ---------- Top panel ----------
    \includegraphics[width=0.78\linewidth]{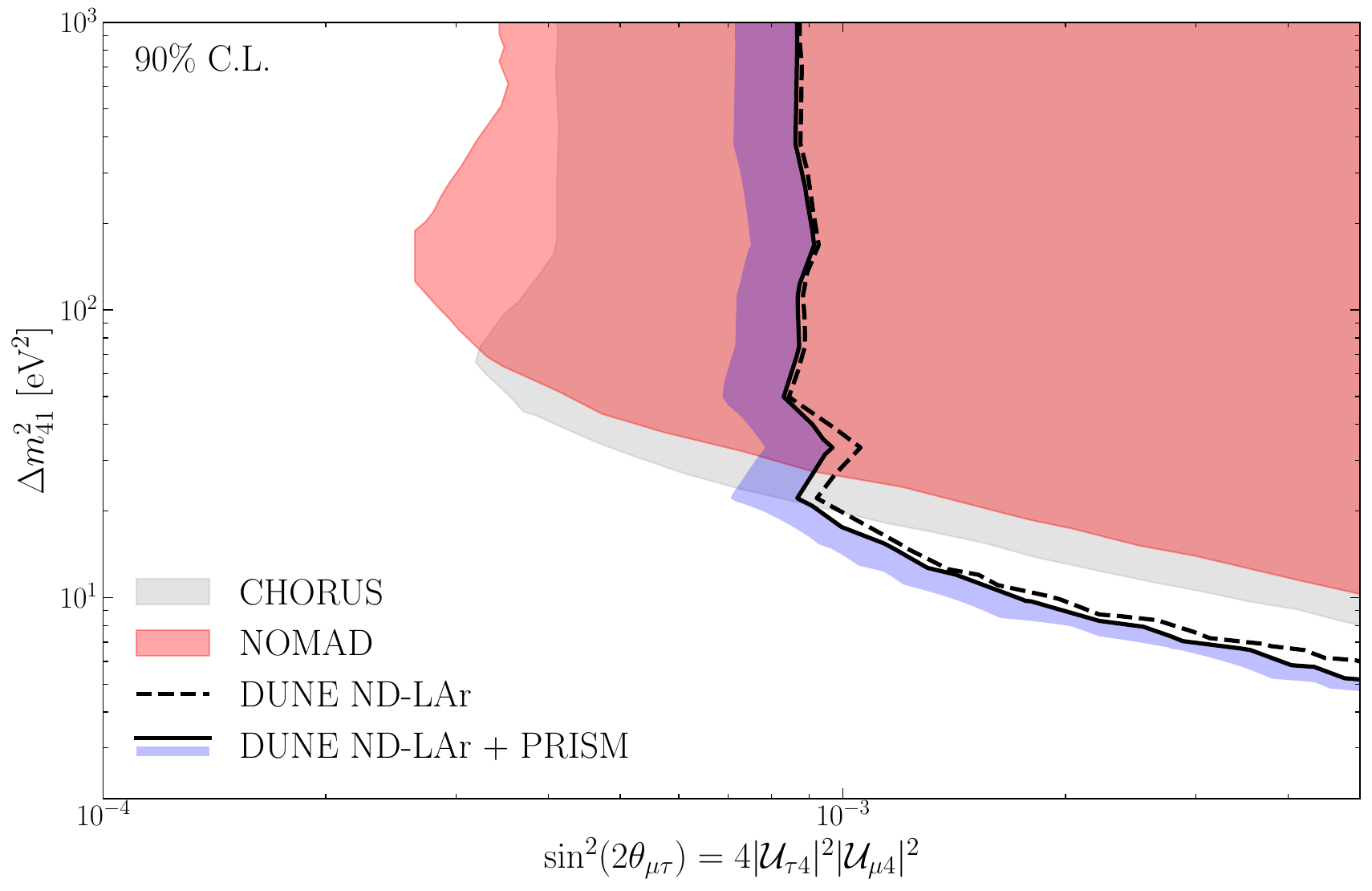}

    \caption{Expected sensitivity to sterile-neutrino appearance channels assuming a $5\%$ shape uncertainty for $P_{\mu \tau}$. The shaded pink and gray areas show current constraints from other experiments at $90\%$ CL~\cite{Tsenov:2009jca, Astier:2001yj}. The solid curve shows the DUNE sensitivity, while the dashed curve shows the sensitivity with DUNE-PRISM. In all cases, the region to the right of the curves is excluded at $90\%$ CL (2 d.o.f.). The blue shaded band highlights the gain in sensitivity obtained from an additional run with a $\tau$-optimized beam.}
    \label{fig:sterile_app_tau}
\end{figure}

Finally, our results for sterile neutrinos in the 3+1 scenario for $\tau$ appearance are presented in Figure~\ref{fig:sterile_app_tau}. We also show, for comparison, the current bounds from NOMAD (red) and CHORUS (grey). In this case, the sensitivity improves upon existing constraints only for mass splittings $\Delta m^2_{41} \lesssim 20~\mathrm{eV}^2$, as in this region the sensitivity of both NOMAD and CHORUS becomes negligible.

In summary, given the current constraints from other neutrino oscillation experiments, $\nu_\tau$ detection in the DUNE near detector complex only leads to a marginal sensitivity improvement over the current bounds in the new physics scenarios under consideration.

\newpage
%\bibliographystyle{JHEP} 
%\bibliography{Refs}
\providecommand{\href}[2]{#2}\begingroup\raggedright\endgroup

\end{document}